 \documentclass[%
 aip,
 prl,%
 amsmath,amssymb,
 reprint,%
twocolumn,
]{revtex4-2}

\usepackage{hyperref} 
\usepackage{graphicx}
\usepackage{dcolumn}
\usepackage{bm}
\usepackage{float}
\usepackage{hyperref}
\usepackage[margin = 1in]{geometry}
\usepackage{lineno}
\usepackage{afterpage}
\usepackage{color}
\usepackage{amsmath}
\usepackage{nccmath} 

\begin{document}

\preprint{APS/123-QED}

\title{Thermal Noise Measurement Below the Standard Quantum Limit}%
\author{Ronald Pagano}
\email{rpagan3@lsu.edu}
\author{Scott Aronson}
\affiliation{Department of Physics \& Astronomy, Louisiana State University, Baton Rouge, LA, 70803}
\author{Torrey Cullen}
\affiliation{California Institute of Technology - The Division of Physics, Mathematics and Astronomy 1200 E California Blvd, Pasadena CA 91125}
\author{Garrett D. Cole}
\affiliation{Thorlabs Crystalline
Solutions, Santa Barbara, CA, USA}
\author{Thomas Corbitt}
\affiliation{Department of Physics \& Astronomy, Louisiana State University, Baton Rouge, LA, 70803}


\date{\today}


\date{\today}

\begin{abstract}
 We present a method characterizing thermal noise in an optical cavity independent from quantum noise despite the thermal noise falling below the quantum noise limit. Using this method, we measured the thermal noise contribution from a GaAs/AlGaAs micro-mirror suspended on a GaAs cantilever microresonator when brought to a cryogenic temperature ($\sim25$ K) and incorporated into a  Fabry–Pérot cavity. An optical spring is formed in this optical cavity. Previously, this setup exploited an optical spring to produce a displacement sensitivity falling 2.8 dB below the free-mass standard quantum limit (SQL), as reported by Cullen et. al \cite{Cullen_subSQL}. Here we use a similar setup to measure thermal noise which fell a maximum of 5 dB below the SQL. This measurement, in turn, allowed for an investigation of quantum noise suppression resulting from the optical spring effect, falling a maximum of 10 dB below the SQL.
\end{abstract}

\maketitle

\section{\label{sec:Introduction}Introduction}
The sensitivity of interferometric measurements is ultimately limited by quantum mechanics and thermodynamics. In interferometric force and displacement measurements, quantum noise manifests itself as uncertainty in the photon number measured by a photodetector (shot noise), and in the quantum radiation pressure applied to the end mirrors of the interferometer. The balance between shot noise (SN) and quantum radiation pressure noise (QRPN) imposes the Standard Quantum Limit (SQL), which can be overcome by clever manipulation of the system. Owing to the inherently small scale at which quantum effects are visible, \textit{classical} effects must be subverted for the sensitivity of a detector to approach the SQL. One formidable source of noise, present in any system, is thermal noise (TN). While reducing TN below the SQL is a noteworthy accomplishment, it produces an interesting quandary: how to characterize the TN at this level independently from quantum noise. This quandary is the subject of the letter.

The SQL can be overcome when quantum correlations are created between the laser light and the motion of the interferometer mirrors. Theoretical work in the 1980s first demonstrated that the SQL was not a fundamental limit \cite{Unruh_NATO1,Unruh_NATO2}. Later work proposed several schemes to inject squeezed light into the LIGO detectors, thus overcoming the SQL \cite{Kimble}. More recently Mason et. al. presented an experimental demonstration of an interferometric displacement measurement reaching a sensitivity 1.5 dB below the SQL \cite{Mason_subSQL}. Later, with the injection of squeezed light, quantum noise in the LIGO detector was suppressed 3 dB below the SQL, first with the implementation of frequency independent squeezed light \cite{Yu_subSQL}, and again after the implementation of frequency dependent squeezed light \cite{jia_conditional_subsql}. In this case, unlike Mason et. al. \cite{Mason_subSQL}, the total sensitivity of the LIGO detectors did not fall below the SQL, largely due to the comparatively high level of thermal noise (TN) produced by the Brownian motion of the TiO\textsubscript{2}-doped Ta\textsubscript{2}O\textsubscript{5}/SiO\textsubscript{2} test mass mirror coatings \cite{LIGO_thermal_noise_msr1, LIGO_thermal_noise_msr2}. 

Here we study \textit{the optical spring effect} as an alternative technique that can allow an interferometer to overcome the free mass SQL. The canonical example of an optical spring is a negatively detuned (blue detuned) Fabry-Pérot cavity formed with a movable mirror at one end \cite{thesis_cripe}. In current ground based gravitational wave detectors (LIGO, Virgo, KAGRA) an optical spring could be formed by detuning the signal recycling cavity. This was demonstrated using the Caltech 40 meter prototype interferometer, although not at the quantum limit \cite{Miyakawa_40m_os}. In quantum limited gravitational wave detectors, the optical spring would reduce the quantum noise floor well below the free mass SQL at frequencies close to the optical spring frequency, and could be used as an alternative or in addition to injecting squeezed light into the interferometer  \cite{Chen_ligo_optical_spring}. Additionally, theoretical and experimental work has shown that the quantum noise floor of ground based detectors could be tuned using an optical spring with forewarning of a gravitational wave from the space based gravitational wave detector LISA \cite{Chen_lisa_forewarnings,Aronson_op_spring_tracking}. While the primary application we consider in this letter is gravitational wave detectors, the optical spring effect could be a useful tool to improve quantum limited interferometers, e.g. optomechanical (OM) sensors, capable of reaching sensitivities below the SQL may offer a means to search for dark matter \cite{Carney_mechanical_dark_matter_sensor, snowmass2021whitepaper}.

The effect of an optical spring on the quantum noise floor was demonstrated experimentally using a quantum noise limited Fabry–Pérot cavity, one end of which was formed with a GaAs/AlGaAs micro-mirror suspended on a GaAs cantilever microresonator. A sensitivity 2.8 dB below the free mass SQL was reached \cite{Cullen_subSQL}. This was made possible by the low TN contribution from the GaAs/AlGaAs mirror coating when compared to other highly reflective mirror coatings. This makes the material a promising candidate for use in future generations of gravitational wave detectors \cite{Cole_10fold_reduction, coal2023_algaas_in_gw_detectors}, and allowed for the observation of optomechanically generated squeezed light at room temperature \cite{nancy_squeeze}, the observation of QRPN in the audio band \cite{Cripe_QRPN}, a demonstration that this QRPN is reduced via squeezed light injection \cite{Yap_qrpn_reduction_w_squeezed_light}, and that it can be suppressed when operating a detuned OM cavity  \cite{Cripe_QBAcancellation}. 

In this letter, we characterize the TN from this same GaAs/AlGaAs microresonator mirror at cryogenic temperatures ($\sim 25$ K) and find that the TN reaches 5 dB below the SQL. The goal of this investigation is twofold: to demonstrate a technique to assess the TN of a quantum limited optical cavity, and in turn to determine the effects of quantum noise free from TN.

\subsection{\label{sec:TN}Thermal Noise}

The TN present in any interferometer is governed by the fluctuation-dissipation theorem. While TN from viscous damping will affect any interferometric measurement left in atmosphere, this source of TN can be eliminated by placing the device in a vacuum, as was done here. Thus, the TN results from the internal damping of the oscillator/mirror coating with a displacement power spectral density (PSD)  \cite{Saulson_thermal_noise, Gonzalez_thermal_noise, Saulson_fundamentals}:
\begin{equation}
    S_\mathrm{th}(\omega)=\sum_{k = 0}^n\frac{4 k_B T \omega_k^2\phi}{m_k\omega[(\omega^2_k-\omega^2)^2+\omega^4_k\phi^2]}.
\end{equation}
Here $T$ is the temperature, $k_B$ is the Boltzmann constant, $\omega$ is the angular frequency, $m_k$, and $\omega_k$ refer to the mass and frequency of the the $k_\mathrm{th}$ order mode. The loss angle $\phi=\frac{1}{Q}$, where $Q$ is the mechanical quality factor of the oscillator. 

Two separate measurements are required in order to characterize the TN spectrum: the quality factor, $Q_k$, and the modal masses of the oscillator. In the experiment, the quality factor was determined to be $25000 \pm 2200$ at cryogenic temperatures via a ring-down measurement. The modal masses were determined from the measured noise spectra. While each modal mass of the mirror, $m_k$, and its corresponding frequency, $\omega_k$, can be predicted with finite element models such as COMSOL (the software used here), these models must be confirmed as they will change if the alignment of the optical beam moves from the center of the mirror. As the alignment cannot be predicted \textit{a priori}, a direct measurement is required.  Characterizing other stationary noise sources such as the residual gas noise \cite{Martynov_cpsd_measurement}, electro-optic noise \cite{Tanioka_electro_optic_AlGaAs}, or the newly discovered birefringence noise \cite{Yu_birefringent_noise} also requires direct measurement. The techniques presented here confirmed the model and enabled us to account for the increased modal masses. They may allow for the characterization of other stationary noise sources in the future. 

\section{Experimental Setup and Procedures}

\begin{figure*}
\includegraphics[width = \textwidth]{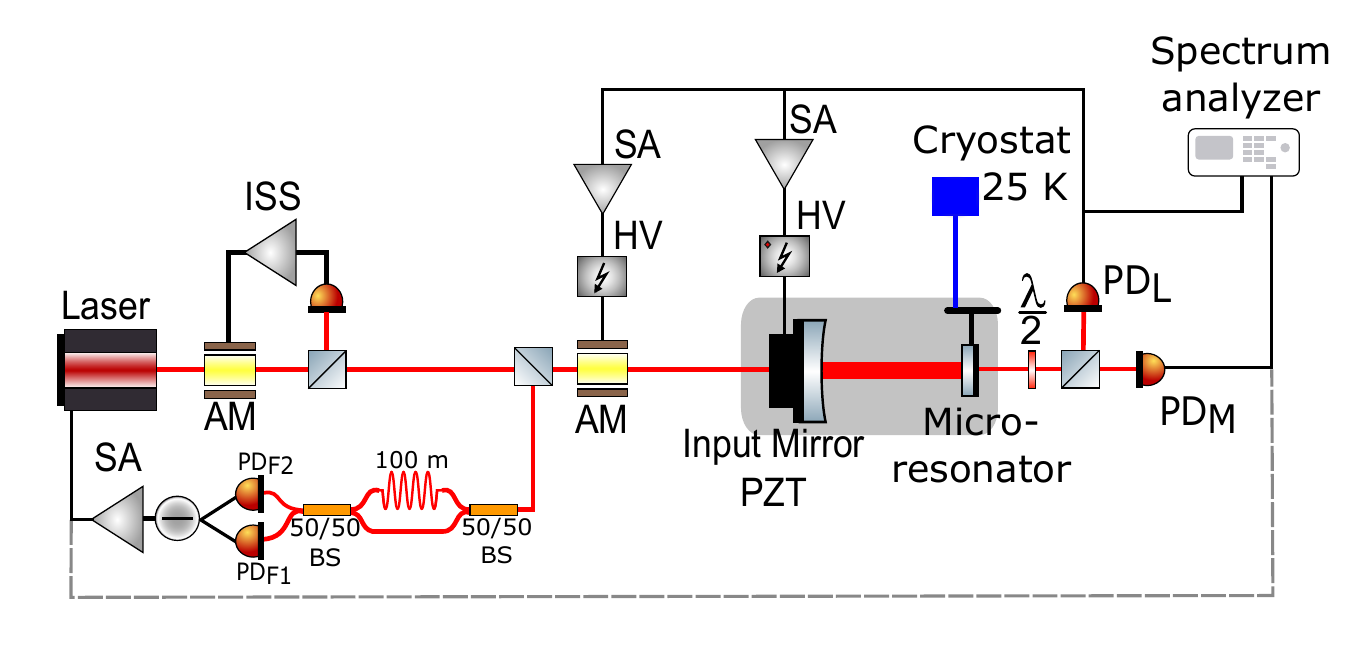}
\caption{\textbf{Overview of the experimental setup.}  
This simplified experimental setup displays the intensity stabilization stage (ISS), the Mach Zehnder interferometer with 100 m delay line to characterize and control the laser frequency noise, and the cryogenically cooled OM cavity formed from a PZT mounted input mirror and GaAs/AlGaAs micro-resonator. A two stage feedback control loop feeds to the PZT input mirror and an amplitude modulator (AM) which stabilizes the OM cavity using a radiation pressure locking scheme \cite{Cripe_RPL}. The feedback actuating on the PZT responds to any slow scale drift in the power circulating in the cavity.}
\label{fig:Setup}
\end{figure*}

Our experimental setup closely resembles the one described in Cullen et. al. \cite{Cullen_subSQL}, consisting of a 1-cm long OM Fabry-Pérot cavity formed from a half inch input mirror with a radius of curvature of $2.5$ cm and the GaAs cantilever with a $70$ $\mathrm{\mu}$m diameter low-noise GaAs/AlGaAs Bragg reflector (discussed above) which serves as the end mirror of the cavity. The suspended micromirror has a mass of $50$ ng and a fundamental resonance frequency of $876$ Hz. The input coupler mount incorporates a piezoelectric crystal that is controlled via a high voltage amplifier, which allows the cavity detuning to be adjusted easily. The OM cavity is housed in a vacuum chamber ($\sim 10^{-8}$ torr) and is cooled cryogenically to approximately $25$ K. The cavity is driven by a 1064 nm Nd:YAG non-planar ring oscillator (NPRO) laser. The laser light is stabilized with an intensity stabilization servo (ISS) before light is picked off and sent to a Mach Zender (MZ) interferometer with a $100$ m delay line in one arm. The output from the interferometer is split onto two photodetectors (PDs), $PD_{F1}$ and $PD_{F2}$, and the differential signal is obtained by subtracting the signal measured on the two PDs. (This balanced photodetector scheme is hereafter referred to as $PD_F$.) This results in a precise measure of the laser frequency noise (LFN) which is sent to a  servo amplifier (SA) where the LFN is suppressed and the frequency is locked to a stable value. Even after the active control of the laser frequency, the LFN is still one of the dominant sources of noise in uncalibrated measurements. This stabilized laser light drives the OM cavity. Downstream from the cavity, the light is split evenly onto two PDs, $PD_L$ and $PD_M$. The signal from $PD_L$ is then used to lock the cavity at a fixed detuning in terms of cavity line-widths ranging from -1.2 to -4.4 using a radiation pressure locking schema \cite{Cripe_RPL}. 

All measurements were taken on a low noise SR785 spectrum analyzer with two inputs which were used to make two separate coincident measurements: first the signals from $PD_L$ and $PD_F$, second the signal from $PD_L$ and $PD_M$. By measuring the LFN we can calibrate our measurements into $\mathrm{m/\sqrt{Hz}}$ and subtract off the remaining LFN. With the second measurement, we are able to calibrate a spectrum free of SN as discussed in Sec. \ref{secTH:QN mitigation}.  While the cavity is cooled to cryogenic temperatures, the cryopump introduces significant vibrations to the micro mirror. Thus we turned off the cryopump at the time of the measurements, and the temperature increased slightly. The temperature was recorded at the end of the measurement.

\subsection{Quantum Noise Mitigation}
\label{secTH:QN mitigation}

If we consider the power fluctuations of the laser light leaving the OM cavity, we can measure this light on two separate PDs by splitting the light with a 50:50 beam splitter (see $PD_\textrm{L}$ and $PD_\textrm{M}$ in Fig.\ref{fig:Setup}). The vacuum fluctuations on each PD is then:
\begin{equation}\label{eq:p_pdL}
     \hat{P}_\mathrm{L,0}=\frac{1}{\sqrt{2}}B   (\hat{v}_\mathrm{0}+\hat{v}_\mathrm{bs}+\hat{v}_\mathrm{cl})+\hat{P}^\mathrm{el}_\mathrm{L},
\end{equation}
\begin{equation}\label{eq:P_pdeqM}
    \hat{P}_\mathrm{M,0}=\frac{1}{\sqrt{2}}B(\hat{v}_\mathrm{0}-\hat{v}_\mathrm{bs}+\hat{v}_\mathrm{cl})+\hat{P}^\mathrm{el}_\mathrm{M}.
\end{equation}
B is the field amplitude; $\hat{v}_\mathrm{0}$ and $\hat{v}_\mathrm{bs}$ are vacuum fluctuations exiting the cavity and coupling into the signal through the unused port of the beamsplitter respectively; $\hat{v}_\mathrm{cl}$ results from any classical noise which most notably includes TN, and $\hat{P}^\mathrm{el}_\mathrm{L/M}$ represents the power fluctuations from uncorrelated electronics noise (e.g. the dark current from the two PDs). Both the SN and QRPN are contained in the vacuum fluctuations of the light. Equations \ref{eq:p_pdL} and \ref{eq:P_pdeqM}  are equivalent to Equation (9) in Martynov et. al. \cite{Martynov_cpsd_measurement}.

The methods we used to mitigate quantum noise and characterize TN are as follows: the SN and uncorrelated electronics noise are canceled by measuring the correlated signal between the two PDs, provided the measurement is averaged over a sufficiently long period of time. The cross power spectral density (CPSD), $S_\mathrm{LM}$ calculated from $P_\mathrm{L,0}$ and $P_\mathrm{M,0}$ gives:
\begin{eqnarray} \label{eqTH:S_LM}
S_\mathrm{LM}&&=S_\mathrm{b,0}-S_\mathrm{b,bs}+S_\mathrm{cl}\nonumber\\
&&=S_\mathrm{rp}+S_\mathrm{cl}.
\end{eqnarray}
$S_\mathrm{b,0}-S_\mathrm{b,bs}=S_\mathrm{rp}$, where $S_\mathrm{rp}$ is the PSD of the QRPN in the measured quadrature. The dominant contribution of classical noise, $S_\mathrm{cl}$, in the final measurement is TN. This analysis resembles that found in \cite{Martynov_cpsd_measurement}, in which the correlated signal of the LIGO detectors was measured, canceling SN, in order to characterize the TN. However, here the measurement is performed using a detuned OM cavity in which an optical spring is formed. The optical spring amplifies the motion of the OM cavity at resonance which is largely due to TN, QRPN, and any displacement signal that we wish to measure, but does not amplify SN \cite{Cullen_subSQL, Aronson_op_spring_tracking, thesis_Pagano}. In order to characterize the sensitivity of the device we must relate our final noise spectrum back to the equivalent motion of the mirror in the free mass regime by dividing the power spectrum by the gain amplifying the signal, $G_\mathrm{os}$:
\begin{equation}
    \frac{S_\mathrm{LM}}{G^2_\mathrm{os}}= S_\mathrm{rp,fm}+S_\mathrm{cl,fm}.
\end{equation}

In our experiment some of the canceled noise is imprinted onto the light by the feedback applied to $PD_\mathrm{L}$, necessary to stabilize the system. When we account for this feedback we find an expression for the cavity signal referred to the free mass:
\begin{equation}
    S_{0}=S_\mathrm{rp,fm}+S_\mathrm{th,fm}+\left(\frac{1}{G_\mathrm{os}}\right)^2\left(S_\mathrm{sn,L}+S_\mathrm{el,L}\right).
    \label{eqTH:S0 cavity spec}
\end{equation}
The cavity signal after the cancellation of SN and uncorrelated electronics noise is:
\begin{equation}
    S_1=S_\mathrm{rp,fm}+S_\mathrm{th,fm}+\frac{G_\mathrm{fb}}{G^2_\mathrm{os}}\left(S_\mathrm{sn,L}+S_\mathrm{el,L}\right).
    \label{eqTH:S1 thermal and qrpn spec}
\end{equation}
$S_\mathrm{sn,L}$ and $S_\mathrm{el,L}$ are the respective SN and electronics noise measured on $PD_\mathrm{L}$, $G_\mathrm{fb}$ is the open loop gain of the feedback. Here we assume the classical noise is dominated by TN ($S_\mathrm{cl,fm}=S_\mathrm{th,fm}$). The feedback does imprint some of the uncorrelated noise on the cavity signal, resulting in an imperfect cancellation of the noise. However, this effect is insignificant in all but one of our measurements presented here, as the amount of noise imprinted on the signal at frequencies close to or below the optical spring (OS) frequency is negligible. At frequencies below the optical spring (OS) frequency the feedback is inactive, and close to the optical spring frequency SN is too small to impact the measurement. Only in the measurement with a 50 kHz OS frequency are any of the frequencies we consider seriously affected by the feedback. A full discussion and derivation of Eq. \ref{eqTH:S0 cavity spec} and \ref{eqTH:S1 thermal and qrpn spec} can be found in the appendix. 

The QRPN is unaffected by the correlation measurement. For this reason the QRPN cancellation, resulting from the optical spring effect first realized in \cite{Cripe_QBAcancellation}, is investigated and compared to the straightforward approach of changing the power of the input laser light. The 
displacement amplitude spectral density of the uncanceled QRPN is given by \cite{Cullen_subSQL}:
\begin{equation}
\sqrt{S_\mathrm{rp,fm} } =\frac{1}{m \omega^{2}} \frac{2 P_\mathrm{c}}{c} \sqrt{\frac{2 h f}{P_\mathrm{i n}}}.
\label{eqTH:qrpn}
\end{equation}
Here $m$, $h$, $f$, and $c$ are the reduced mass of the oscillator, Planck's constant, the frequency of the light, and the speed of light respectively. $P_\mathrm{in}$ and $P_\mathrm{c}$ are the input power and the circulating power of the light. We see the QRPN is reduced by increasing $P_\mathrm{in}$, the power of the light pumping the cavity, if $P_\mathrm{c}$, the light circulating inside the cavity is held constant. This is achieved by changing the input power in conjunction with the detuning; the power of the light circulating in the cavity is given by 
\begin{equation}
    P_\mathrm{c}= \frac{P_\mathrm{0}}{1+\delta^2},
\end{equation}
where $\delta$ is the fraction of the cavity linewidth by which the cavity is detuned. $P_0$ is the maximum circulating power which is directly related to the input power (if the only loss is the transmission through one mirror $T$, $P_0=4P_\mathrm{in}/T$). The QRPN can also be lowered by decreasing the power of the light circulating in the cavity if the input power is held constant. When we include the QRPN cancellation resulting from the optical spring effect, our numeric model \cite{Corbitt_mathematical} shows that the measured QRPN is suppressed by a factor $ \omega/\omega_\mathrm{os}$ at frequencies below the OS frequency ($\omega$ is the measured frequency and $\omega_{os}$ is the frequency of the optical spring). This QRPN cancellation scheme is similar to the variational readout scheme utilizing correlations between SN and QRPN, reducing the quantum noise limit \cite{vyatchanin_quantum_variation_measurement, Kimble, chen_qnd_for_gw_detectors}. In this experiment we demonstrate two of these methods. We changed the input power of the laser light pumping the OM cavity while the light circulating inside the cavity was held constant, resulting in a change in OS frequency. We show that the QRPN was suppressed below the OS frequency, which is most obvious in the 69 kHz OS measurement (see Figure \ref{figTH:qrpn suppression}). However, this suppression is eclipsed by simply increasing the input power while holding the circulating power fixed. This can be seen by comparing the 69 kHz OS data to the lower OS frequency data sets corresponding to higher input power. The latter shows a much smaller contribution from QRPN.

While we wish to mitigate quantum noise to characterize TN, these techniques could improve the sensitivity of interferometric detectors searching for non-transient signals as the cross correlation measurements must be averaged. These include gravitational waves produced by known pulsars such as the crab pulsar, which current interferometric gravitational wave detectors cannot detect \cite{LIGO_GW_from_pulsars, LIGO_crab_pulsar}. Evidence for the stochastic gravitational wave background produced by binary black hole systems has been gathered by observing pulsar timing arrays \cite{NANOGrav_15yr} but has not been detected by any interferometric gravitational wave detector. Finally, dark matter candidates could leave a stochastic signal on mechanical sensors \cite{gardner_stochastic_quantum_limit}.

\section{Results and discussion}

In order to obtain the best estimate for the TN free from quantum noise and study the changing quantum noise, several measurements were made at different OS frequencies. Four measurements are displayed in Figs. \ref{figTH:cavity and TH specs}, \ref{figTH:qrpn suppression}, and \ref{figTH:quantum noise specs} taken with input powers of 0.7, 2.5, 5, and 7 mW and detunings (given in terms of cavity linewidths) of -1.2, -2.5, -3.5, and -4.4. These parameters resulted in measured OS frequencies of 69, 63, 58, and 50 kHz (from top to bottom in all three figures). When the OS frequency and input power changed, the quantum noise changed in response, while the effect of Brownian TN was unaffected. All measurements were taken at $\sim 25$ K.

The uncertainties of the measured and modeled spectrum were accounted for as follows. The primary source of uncertainty of $\sim 1\%$ identified in the $S_0$ spectrum results from the uncertainty in the calibration from  $\mathrm{V^2/Hz}$ into $\mathrm{m^2/Hz}$, just as in \cite{Cullen_subSQL}. Additional variation in $S_1$ spectra resulted from variations in the measured SN and uncorrelated electronics noise. A $10\%$ uncertainty in the mass of the mirror, an $8\%$ uncertainty in the temperature,  and a $9\%$ uncertainty in the power of the light pumping the cavity all result in uncertainty in our modeled noise spectrum. 

\subsection{Thermal Noise Measurement}\label{secTH:Th_measure}

\begin{figure*}
    \includegraphics[width = \textwidth]{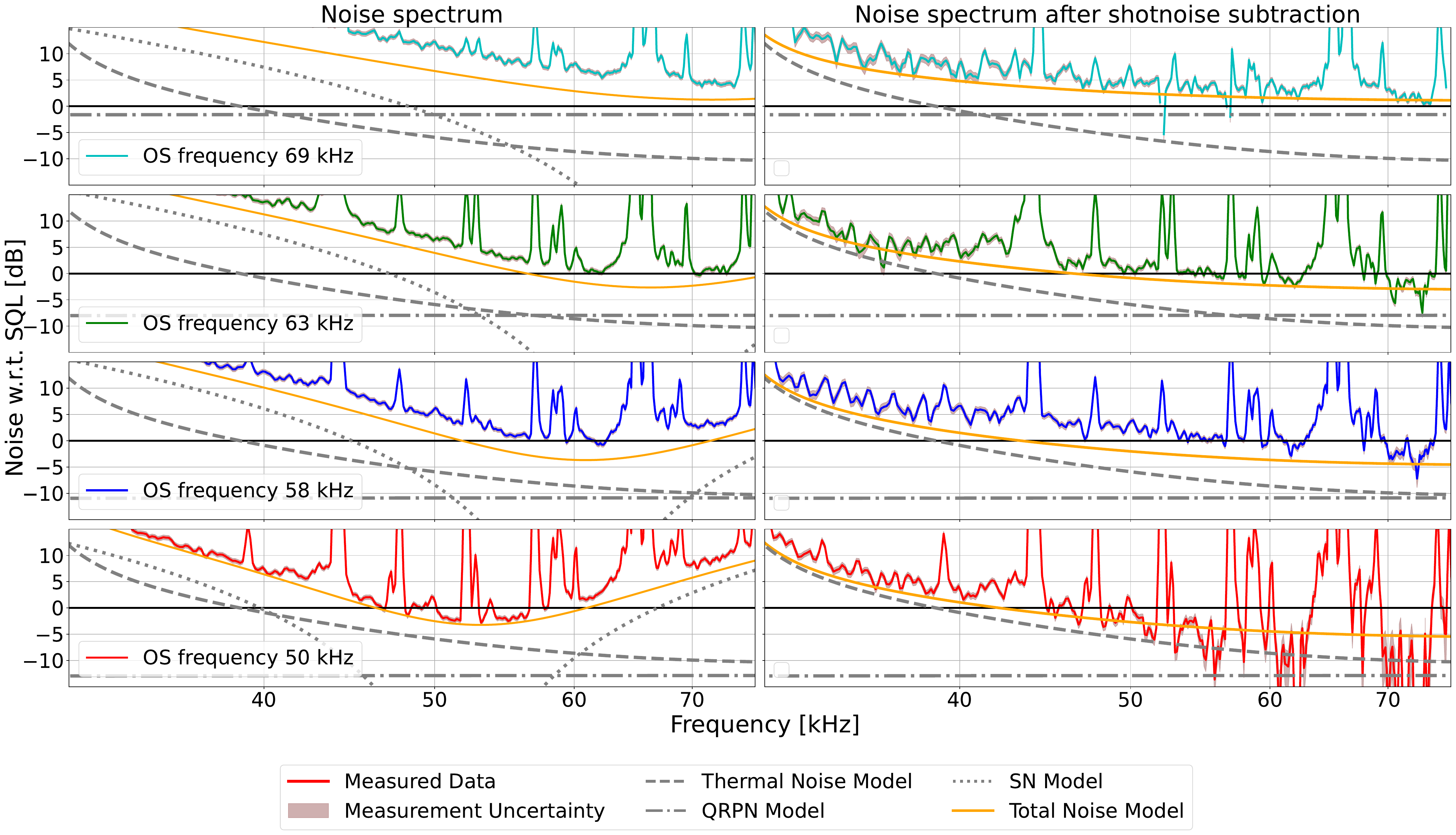}
    \caption{Measured spectrum normalized to the SQL before and after SN is removed: the data presented here was used to assess the TN at sensitivities falling below the SQL. The plots in the first column present the measured ASD before the SN is removed, normalized to the SQL. The second column presents the ASD of the same data after SN and uncorrelated electronics noise are removed. The modeled TN, QRPN, and SN are displayed separately, along with the sum of these noise sources (total modeled noise) and the measured spectrum in each plot. All measurements show clear evidence that the SN is removed in the second column. The 50 kHz OS frequency spectrum falls 2.2 dB below the SQL before the uncorrelated noise is removed and $\sim 2$ dB below the SQL after the SN is removed at frequencies below the OS frequency.}
    \label{figTH:cavity and TH specs}
\end{figure*}

Fig. \ref{figTH:cavity and TH specs} displays measurements in two columns: the first presents the amplitude spectral density (ASD) in units of $\mathrm{m/\sqrt{Hz}}$ before the SN is removed, normalized to the free mass SQL ($\sqrt{S_0/S_\mathrm{SQL}}$, where  $S_0$ is given by Eq. \ref{eqTH:S0 cavity spec}, and $S_\mathrm{SQL}=\frac{2\hbar}{m\omega^2}$). The second column presents the ASD after SN and the uncorrelated electronics noise is removed, also normalized to the SQL ($\sqrt{S_1/S_\mathrm{SQL}}$, with $S_1$ given by Eq. \ref{eqTH:S1 thermal and qrpn spec}). All datasets display unmodeled lines throughout the measurements, which is most likely the result of electronics noise. There is also some separation between the modeled and measured noise, which is more pronounced before the uncorrelated noise is removed, suggesting that not all optical losses present in the cavity are accounted for, or that uncorrelated electronics noise is not negligible. All measurements show clear evidence that SN is removed in the second column. The 50 kHz OS frequency spectrum falls 2.2 dB below the SQL before the uncorrelated noise is removed. Much of the frequency range considered here exceeds the 50 kHz OS frequency present in this measurement, meaning that the cross correlation measurement is unreliable at these frequencies (see discussion in Sec. \ref{secTH:QN mitigation} and Appendix \ref{secApp:CPSD noise subtract}). However, we do see clear evidence of the $S_1$ spectrum falling $\sim 2$ dB  below the SQL at frequencies below the OS frequency.

\begin{figure}[t!]
    \includegraphics[width=8cm] {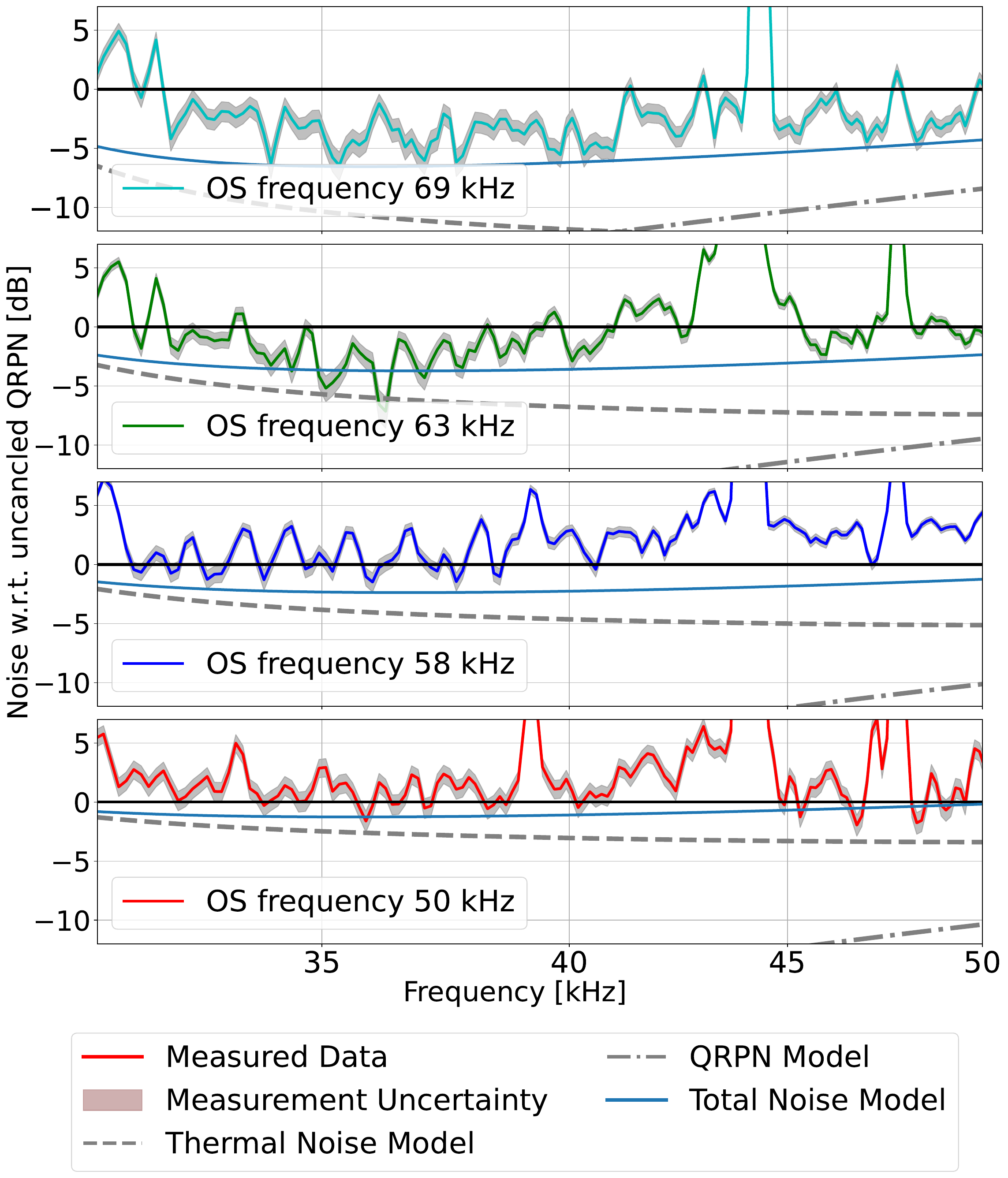}
    \caption[Measured Spectra Normalized to Modeled SN and Unsuppressed QRPN]{Measured spectra normalized to modeled SN and unsuppressed QRPN: the 69 kHz OS dataset (top plot) shows suppression of the QRPN over a large frequency range, and the 63 kHz OS dataset shows suppression in a smaller frequency range. The 58 and 50 kHz OS frequency datasets show little effect from the QRPN suppression, although the measured QRPN is at a lower level than in the 69 and 63 kHz OS datasets. All plots display the modeled TN and QRPN, the sum of these noise sources, and the measured noise. }
    \label{figTH:qrpn suppression}
\end{figure}

\begin{figure}[h!]
    \includegraphics[width=8cm] {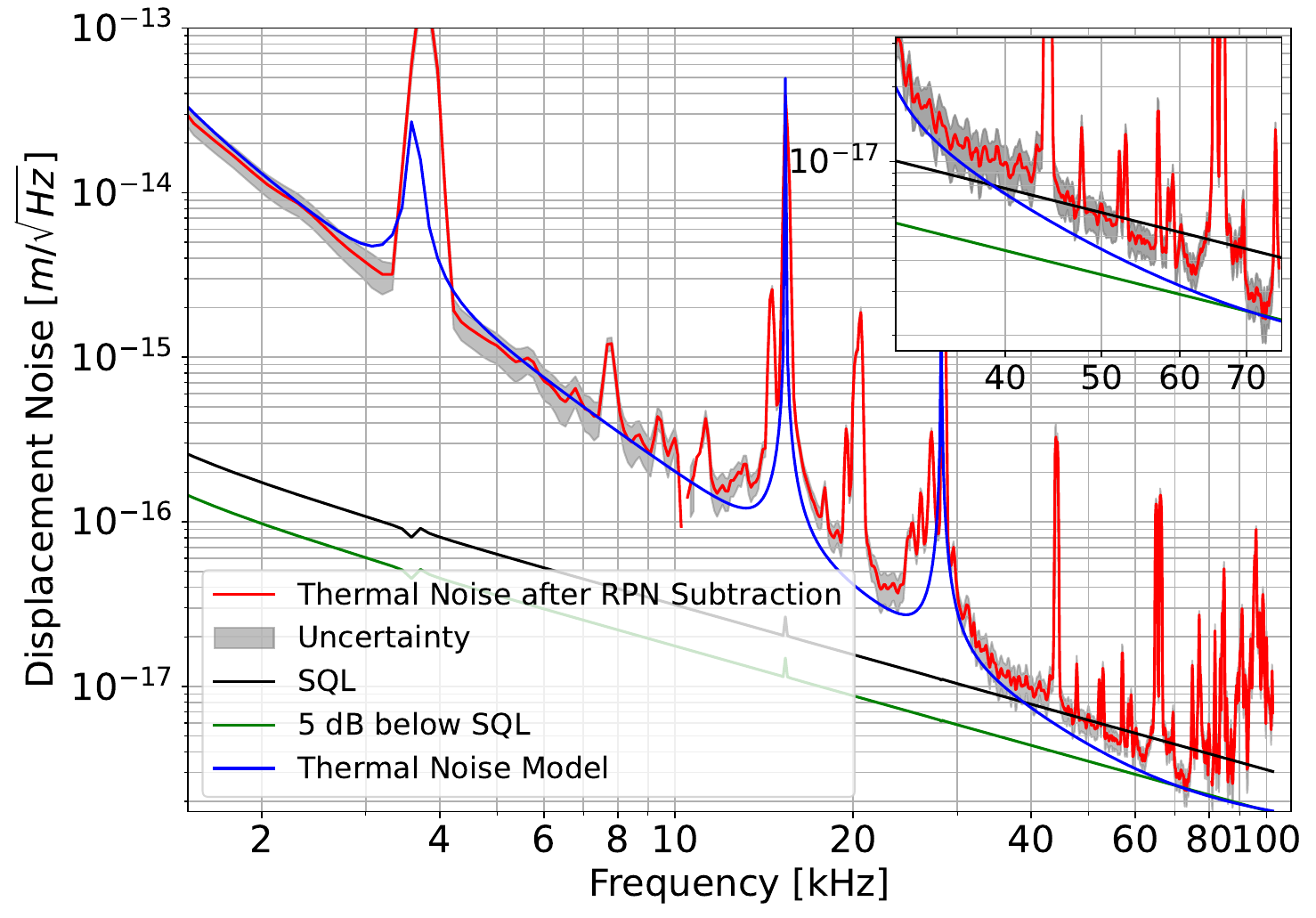}
    \caption[Plot Displaying the Averaged TN Measurement]{Plot displaying the average TN measurement free from quantum noise. Here the 69, 63, and 58 kHz OS frequency datasets, along with an additional dataset with an OS frequency of 67 kHz are averaged together after the modeled QRPN is subtracted. The 50 kHz OS dataset is excluded as a large portion of the data is above the 50 kHz optical spring frequency, making the measurement unreliable at these frequencies. Here we see good broadband agreement between the model and the measurement. Noise falling below the SQL by a maximum of 5 dB is measured. The model and measurement of TN, along with the SQL and a line marking 5 dB below the SQL, are plotted.}
    \label{figTH:Avg Th noise}
\end{figure}

In all these measurements QRPN remains close to the TN floor. Here we see that the largest contribution of QRPN is in the 69 kHz OS dataset, and the smallest in the 50 kHz OS dataset. The suppression of QRPN was investigated by comparing the modeled unsuppressed QRPN and TN to the measured $S_1$ spectrum as displayed in Fig. \ref{figTH:qrpn suppression}. Here we see suppression of the QRPN over a large frequency range in the 69 kHz OS dataset (top plot), and in a smaller frequency range in the 63 kHz OS dataset (second plot from the top). However, the 58 and 50 kHz OS datasets (bottom two plots) show little effect from the QRPN suppression, although the measured QRPN is at a lower level than the 69 and 63 kHz OS datasets. 
Because the modeled QRPN remains close to the TN floor in all measurements, we subtracted off the modeled QRPN to obtain the best assessment of TN free from quantum noise. Here the 50 kHz OS dataset is excluded from the average TN, while a fourth dataset, measured with an input power of 1.5 mW and a detuning of -1.9, resulting in an OS frequency of 67 kHz, is included in the average. An ultimate sensitivity of 5 dB below the SQL is observed in the TN measurement, presented in Fig. \ref{figTH:Avg Th noise}. The high frequency noise above 70 kHz has been measured consistently in electronics used in the experiment. Other lines have not been linked as clearly to electronics in the setup, although the modeled modes of our cantilever mirror do not match many of these lines. We see only a small disagreement between our modeled and measured broadband TN floor. Thus, we find strong evidence that our model accurately represents our system. 

\subsection{Quantum Noise Measurement}
\begin{figure}[h!]
    \includegraphics[width=8cm]{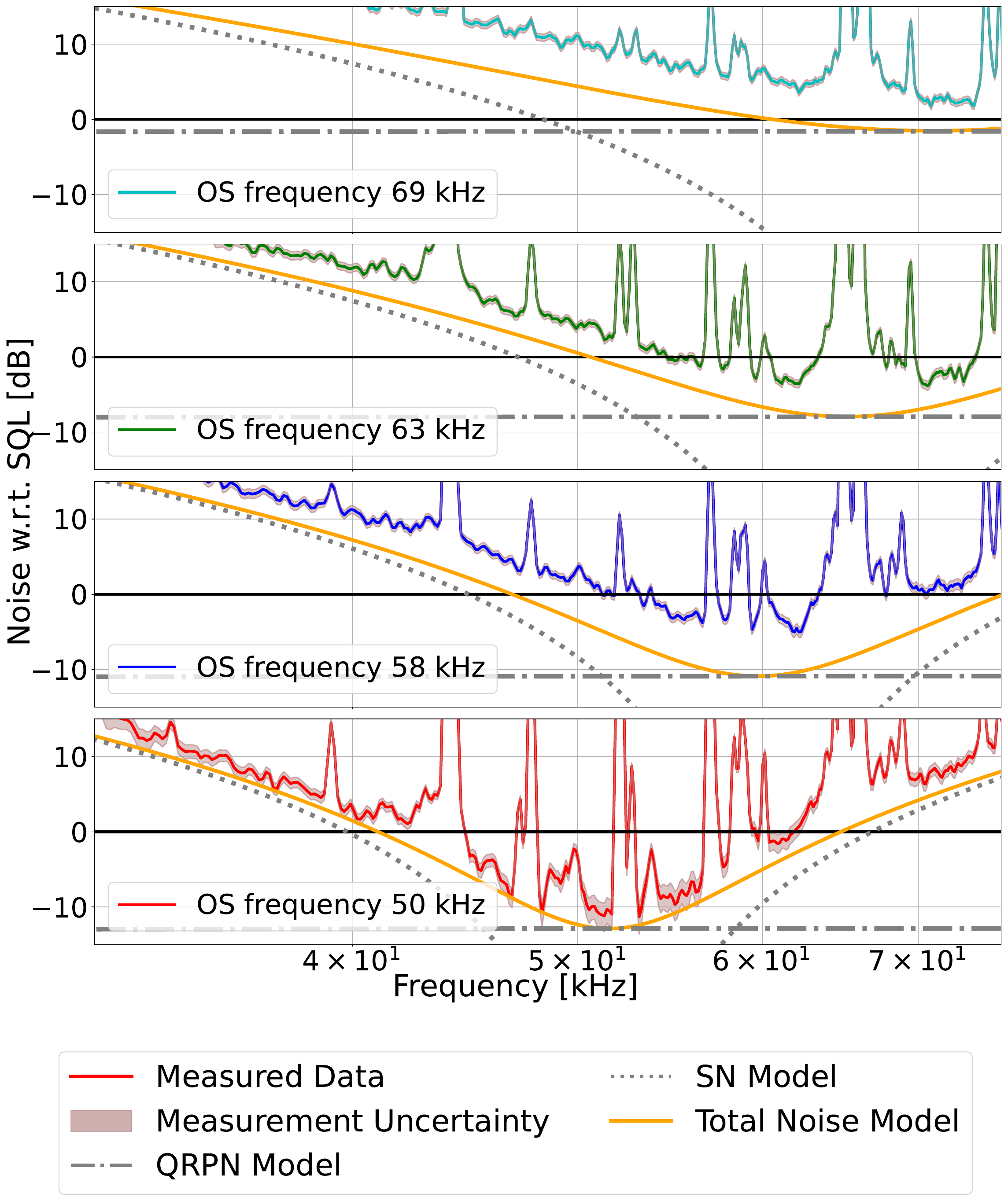}
    \caption[Plot Displaying Measured Quantum Noise Normalized to the SQL]{Plot displaying measured quantum noise spectra normalized to the SQL after the modeled TN is subtracted. Here we see all but the 69 kHz OS measurement falling below the SQL, with a maximum suppression of $\sim10$ dB below the free mass SQL observed in the 50 kHz measurement.}
    \label{figTH:quantum noise specs}
\end{figure}

Given that Fig. \ref{figTH:Avg Th noise}, confirms our modeled TN, this model was subtracted from our measured spectra presented in the first column of Fig. \ref{figTH:cavity and TH specs}. This produced an upper bound for the quantum noise in these measurements. Fig. \ref{figTH:quantum noise specs} displays the spectra. Here we see all but the 69 kHz OS measurement falling below the SQL, with a maximum suppression of $\sim10$ dB below the free mass SQL, observed in the 50 kHz OS dataset. This marks a significant improvement over the result reported in \cite{Yu_subSQL}, and \cite{Mason_subSQL}.

\section{Discussion and Outlook}
Developing interferometers capable of performing a displacement measurement with a sensitivity that falls below the SQL has been a long-standing goal in the quantum optics community.  In recent years this has been achieved, although significant challenges persist, notably the barrier imposed by TN. The use of GaAs/AlGaAs mirror coatings is one means of reducing the limitation imposed by TN. Our GaAs/AlGaAS microresonator allows for sensitivities well below the SQL to be reached, as we utilized the optical spring effect, greatly reducing quantum noise below the SQL.

Our primary motivation was to improve interferometric gravitational wave detectors. This work emphasizes the potential usefulness of the optical spring effect in mitigating quantum noise in such detectors. The LIGO detectors are currently limited by TN. An effort to implement GaAs/AlGaAs mirror coatings onto the LIGO test masses is currently under way \cite{coal2023_algaas_in_gw_detectors}. The degree by which TN can be reduced in comparison to quantum noise will ultimately determine how useful the implementation of an optical spring in the detector can be. 

Beyond gravitational wave detectors, the use of a GaAs/AlGaAs microresonator in our Fabry Pérot cavity, in combination with the optical spring effect is a proven means of overcoming the free mass SQL. Such an OM sensor may be useful for applications in which the sensitivity of the device must reach below the SQL. 

\begin{acknowledgments}
R. Pagano, S.A, H.C., and T. Corbitt are supported by the National Science Foundation grants PHY-2110455. T.C. is supported by in part by the Heising-Simons Foundation through grant 2022-3341. A portion of this work was performed in the UCSB Nanofabrication Facility, an open access laboratory. 

\end{acknowledgments}

\appendix

\section{Calibration} 
\label{secApp:CPSD noise subtract}
In order to remove the LFN and complete the calibration into $\mathrm{m/\sqrt{Hz}}$, a cross power spectral density (CPSD) between $PD_\mathrm{F}$ and one of the PDs monitoring the cavity signal was recorded along with the PSD of both $PD_\mathrm{L}$ and $PD_\mathrm{F}$. While either $PD_\mathrm{L}$ or $PD_\mathrm{M}$ could be used, $PD_\mathrm{L}$ was used here. The derivation of a conversion factor, used to calibrate the spectra into the appropriate units is documented in \cite{Cullen_subSQL, thesis_Pagano}. 
In order to remove the SN, a second CPSD and the corresponding PSDs are measured between $PD_\mathrm{L}$ and $PD_\mathrm{M}$. To give a full account of the calculations performed on this data, we first consider the power fluctuations measured on $PD_\mathrm{F}$, given by: 
\begin{equation}
    \hat{P}_\mathrm{F} = A(\hat{v}_\mathrm{v,f}+\hat{v}_\mathrm{f})+\hat{P}^\mathrm{el}_\mathrm{F}
    \label{eqApp:p_pdF}
\end{equation}
Where $A$ is the field amplitude, $\hat{v}_\mathrm{v,f}$ is the result of vacuum fluctuations measured on $PD_\mathrm{F}$, and $\hat{v}_\mathrm{f}$ is the result of the LFN. When the feedback gain applied to $PD_\mathrm{L}$ is accounted for, the power fluctuations measured on $PD_\mathrm{L}$ become:


\begin{align}
\hat{P}_\mathrm{L}&= \hat{P}_\mathrm{L,0}-G_\mathrm{fb}\hat{P}_\mathrm{L} \label{eqApp:p_pdL} \\ 
&=\frac{1}{1+G_\mathrm{fb}}\left[\frac{1}{\sqrt{2}}B(\hat{v}_\mathrm{0}+\hat{v}_\mathrm{bs}+\hat{v}_\mathrm{cl})+\hat{P}^\mathrm{el}_\mathrm{L}\right].\nonumber
\end{align}

Here $\hat{P}_\mathrm{L,0}$ represents the power fluctuations measured on the respective PDs before the feedback is applied, $G_\mathrm{fb}$ is the open loop gain of the feedback. In the second line B is the field amplitude; $\hat{v}_\mathrm{0}$ and $\hat{v}_\mathrm{bs}$ are vacuum fluctuations exiting the cavity and coupling into the signal through the unused port of the beamsplitter respectively; $\hat{v}_\mathrm{cl}$ results from any classical noise which most notably includes TN, and $\hat{P}^\mathrm{el}_\mathrm{L/M}$ represents the power fluctuations from uncorrelated electronics noise (e.g. the dark current from the two PDs) just as defined in Eq. \ref{eq:p_pdL}.

The resulting power spectrum measured on $PD_\mathrm{F}$ is
\begin{equation}
S_\mathrm{FF}=S_\mathrm{f,fm}+S_\mathrm{sn,F}+S_\mathrm{el,F},
    \label{eqApp:s_ff}
\end{equation}

and on $PD_\mathrm{L}$ is

%

\begin{small} 
\begin{align}
S_\mathrm{LL}&=\left(\frac{1}{1+G_\mathrm{fb}}\right)^2\left(S_\mathrm{0}+S_\mathrm{bs}+S_\mathrm{cl}+S_\mathrm{el,L}\right) \label{eqApp:s_ll1} \\
&=\left(\frac{1}{1+G_\mathrm{fb}}\right)^2 \left(S_\mathrm{rp}+S_\mathrm{sn,L}+S_\mathrm{f}+S_\mathrm{th}+S_\mathrm{el,L}\right).\nonumber
\end{align}
\end{small}
Each element of the Eqs. \ref{eqApp:s_ff} and \ref{eqApp:s_ll1} comes from a corresponding element of the Eqs. \ref{eqApp:p_pdF} and \ref{eqApp:p_pdL}. $S_\mathrm{f,fm}$ is the LFN measured after the MZ interferometer, before the optical spring gain affects the noise source; $S_\mathrm{sn,F}$ is the SN and $S_\mathrm{el,F}$ is the electronics noise measured on $PD_\mathrm{F}$; $S_\mathrm{sn,L}$ and $S_\mathrm{el,L}$ are the respective SN and electronics noise measured on $PD_\mathrm{L}$ before the feedback is applied to the photodetector; last, $S_\mathrm{rp}$ is the noise resulting from the QRPN and $S_\mathrm{cl}= S_\mathrm{f}+S_\mathrm{th}$ is the classical noise, with contributions from the LFN ($S_\mathrm{f}$) and TN ($S_\mathrm{th}$). These spectra represent the noise after it is amplified by the optical spring gain $G_\mathrm{os}$. Expressing the power spectrum explicitly in terms of $G_\mathrm{os}$,


\begin{align}
S_\mathrm{LL}&=\left(\frac{G_\mathrm{os}}{1+G_\mathrm{fb}}\right)^2\left(S_\mathrm{rp,fm}+S_\mathrm{f,fm}+S_\mathrm{th,fm}\right) \nonumber \\
&+\left(\frac{1}{1+G_\mathrm{fb}}\right)^2\left(S_\mathrm{sn,L}+S_\mathrm{el,L}\right).\label{eqn:SLL}
\end{align}

$S_\mathrm{rp,fm}$, $S_\mathrm{f,fm}$, and $S_\mathrm{th,fm}$ are the fractional power spectra refereed back to the free mass regime. $S_\mathrm{el,F}$ and $S_\mathrm{el,L}$ are the contributions of electronics noise on the two PDs. Finally, $S_\mathrm{sn,F}$ is the power spectrum resulting from SN measured on $PD_\mathrm{F}$, and $S_\mathrm{sn,L}$ is the power spectrum resulting from SN measured on $PD_\mathrm{L}$ before the feedback is applied to the system. The resulting CPSD is then:

\begin{equation}
    S_\mathrm{FL}=\frac{G_\mathrm{os}}{1+G_\mathrm{fb}}S_\mathrm{f,fm}.
\end{equation}
Consider the coherence between the signals measured on $PD_\mathrm{L}$ and $PD_\mathrm{M}$:
\begin{equation}
    C=\frac{\left|\langle S_\mathrm{FL}\rangle\right|^2} {S_\mathrm{FF}S_\mathrm{LL}}.
    \label{eqApp:C def}
\end{equation}
Because the power of the light falling on $PD_\mathrm{F}$ is high, the SN contribution $S_\mathrm{sn,F}$, along with any contribution from $S_\mathrm{el,F}$, are negligible. For this reason the coherence is approximately given by:

\begin{small}
\begin{equation}
    C\approx \frac{S^2_\mathrm{f,fm}}{\left[\left(S_\mathrm{rp,fm}+S_\mathrm{f,fm}+S_\mathrm{th,fm}\right)+\frac{1}{G_\mathrm{op}}^2\left(S_\mathrm{sn,L}+S_\mathrm{el,L}\right)\right]S_\mathrm{f,fm}}.
\end{equation}
\end{small}

The noise spectrum free from TN in units of $\mathrm{m^2/Hz}$ is then given by:

\begin{small}
\begin{eqnarray}
    S_{0}&&=K^2_\mathrm{m}S_\mathrm{FF}\left(\frac{1-C}{C}\right) \label{eqApp:S0 cavity signal}\\
    &&=S_\mathrm{rp,fm}+S_\mathrm{th,fm}+\left(\frac{1}{G_\mathrm{op}}\right)^2\left(S_\mathrm{sn,L}+S_\mathrm{el,L}\right).\nonumber
\end{eqnarray}
\end{small}

Here, $K_\mathrm{m}$ is the conversion factor between $V/\sqrt{Hz}$ and $m/\sqrt{Hz}$. We name this spectrum $S_0$, indicating that it contains all fundamental sources of noise along with some contribution of electronics noise. The first column of Fig. \ref{figTH:cavity and TH specs} presents this spectrum normalized to the SQL, $\sqrt{S_0/S_\mathrm{SQL}}$. 

In order to obtain a measurement free from LFN and SN, a second measurement must be take on $PD_\mathrm{M}$. Here the light is split evenly between the two PDs after the cavity. The power fluctuations measured on $PD_\mathrm{M}$ are given by 
\begin{eqnarray}
\hat{P}_\mathrm{M}&&=\hat{P}_\mathrm{M,0}-G_\mathrm{fb}\hat{P}_\mathrm{L} \nonumber  \\
&&=\frac{1}{\sqrt{2}}B(\hat{v}_\mathrm{0}-\hat{v}_\mathrm{bs}+\hat{v}_\mathrm{cl})+\hat{P}^\mathrm{el}_\mathrm{M} \\
&&-\frac{G_\mathrm{fb}}{1+G_\mathrm{fb}}\left[\frac{1}{\sqrt{2}}iB^*(\hat{v}_\mathrm{0}+\hat{v}_\mathrm{bs}+\hat{v}_\mathrm{cl})+\hat{P}^\mathrm{el}_\mathrm{L}\right].\nonumber
\end{eqnarray}

The CPSD between $PD_\mathrm{L}$ and $PD_\mathrm{M}$ is
\begin{eqnarray}
    S_\mathrm{LM}&&=\left(\frac{1}{1+G_\mathrm{fb}}\right)^2(S_\mathrm{0}-S_\mathrm{b}+S_\mathrm{cl})\\ \nonumber
    &&-2\frac{G_\mathrm{fb}}{(1+G_\mathrm{fb})^2}S_\mathrm{b}-\frac{G_\mathrm{fb}}{(1+G_\mathrm{fb})^2}S_\mathrm{el,L} \nonumber \nonumber\\
&&=  S_1+S_\mathrm{exs}.
\end{eqnarray}
which we rewrite as
\begin{equation}
    S_\mathrm{LM}= 
    \left(\frac{G_\mathrm{os}}{1+G_\mathrm{fb}}\right)^2\left(S_\mathrm{rp,fm}+S_\mathrm{f,fm}+S_\mathrm{th,fm}\right)-S_\mathrm{exs},
    \label{eqApp:S_LM}
\end{equation}
with
\begin{equation}
    S_\mathrm{exs}=\frac{G_\mathrm{fb}}{(1+G_\mathrm{fb})^2}S_\mathrm{sn,L}+\frac{G_\mathrm{fb}}{(1+G_\mathrm{fb})^2}S_\mathrm{el,L},
\end{equation}
$S_0-S_\mathrm{b}=G_\mathrm{os}S_\mathrm{rp,fm},$ and $2S_\mathrm{b}=S_\mathrm{sn,L}$. 

We can then arrive at an upper bound of the shotnoise present in $S_0$:

\begin{small}
\begin{eqnarray}
\left(S_{\mathrm{LL}}-S_\mathrm{LM}\right)\times\left(\frac{1+G_\mathrm{fb}}{G_\mathrm{os}}\right)^2 &&=\left(\frac{1}{G_\mathrm{op}}\right)^2\left(S_\mathrm{sn,L}+S_\mathrm{el,L}\right)\nonumber \\ &&+\left(\frac{1+G_\mathrm{fb}}{G_\mathrm{os}}\right)^2\times S_\mathrm{exs}.
\label{eqAPP:sn aprox}
\end{eqnarray}
\end{small}

Note, to evaluate equation Eq. \ref{eqAPP:sn aprox}, we use 
\begin{equation}
    \left(\frac{G_\mathrm{os}}{1+G_\mathrm{fb}}\right)=\frac{S_\mathrm{FL}}{S_\mathrm{FF}}.
\end{equation}
Finally subtracting Eq. \ref{eqAPP:sn aprox} from Eq. \ref{eqApp:S0 cavity signal} we find the spectra, calibrated into $m^2/Hz$:
\begin{equation}
      S_1=S_\mathrm{rp,fm}+S_\mathrm{th,fm}-\frac{G_\mathrm{fb}}{G^2_\mathrm{os}}(S_\mathrm{sn,L}+S_\mathrm{el,L}).
\end{equation}
Here we see that our meaurement is not exact by the quantity $ \frac{G_\mathrm{fb}}{G^2_\mathrm{os}}(S_\mathrm{sn,L}+S_\mathrm{el,L})$, adding some uncertainty to the measurement. However this quantity is negligible close to the optical spring frequency: $G_\mathrm{fb}/G^2_\mathrm{os}<<0$.


\vspace{2em} 
\hrule
\vspace{1em} 
\bibliography{references}

\begin{thebibliography}{37}%
\makeatletter
\providecommand \@ifxundefined [1]{%
 \@ifx{#1\undefined}
}%
\providecommand \@ifnum [1]{%
 \ifnum #1\expandafter \@firstoftwo
 \else \expandafter \@secondoftwo
 \fi
}%
\providecommand \@ifx [1]{%
 \ifx #1\expandafter \@firstoftwo
 \else \expandafter \@secondoftwo
 \fi
}%
\providecommand \natexlab [1]{#1}%
\providecommand \enquote  [1]{``#1''}%
\providecommand \bibnamefont  [1]{#1}%
\providecommand \bibfnamefont [1]{#1}%
\providecommand \citenamefont [1]{#1}%
\providecommand \href@noop [0]{\@secondoftwo}%
\providecommand \href [0]{\begingroup \@sanitize@url \@href}%
\providecommand \@href[1]{\@@startlink{#1}\@@href}%
\providecommand \@@href[1]{\endgroup#1\@@endlink}%
\providecommand \@sanitize@url [0]{\catcode `\\12\catcode `\$12\catcode `\&12\catcode `\#12\catcode `\^12\catcode `\_12\catcode `\%12\relax}%
\providecommand \@@startlink[1]{}%
\providecommand \@@endlink[0]{}%
\providecommand \url  [0]{\begingroup\@sanitize@url \@url }%
\providecommand \@url [1]{\endgroup\@href {#1}{\urlprefix }}%
\providecommand \urlprefix  [0]{URL }%
\providecommand \Eprint [0]{\href }%
\providecommand \doibase [0]{https://doi.org/}%
\providecommand \selectlanguage [0]{\@gobble}%
\providecommand \bibinfo  [0]{\@secondoftwo}%
\providecommand \bibfield  [0]{\@secondoftwo}%
\providecommand \translation [1]{[#1]}%
\providecommand \BibitemOpen [0]{}%
\providecommand \bibitemStop [0]{}%
\providecommand \bibitemNoStop [0]{.\EOS\space}%
\providecommand \EOS [0]{\spacefactor3000\relax}%
\providecommand \BibitemShut  [1]{\csname bibitem#1\endcsname}%
\let\auto@bib@innerbib\@empty
\bibitem [{\citenamefont {Cullen}\ \emph {et~al.}(2024)\citenamefont {Cullen}, \citenamefont {Pagano}, \citenamefont {Aronson}, \citenamefont {Cripe}, \citenamefont {Sharif}, \citenamefont {Lollie}, \citenamefont {Cain}, \citenamefont {Heu}, \citenamefont {Follman}, \citenamefont {Cole}, \citenamefont {Aggarwal},\ and\ \citenamefont {Corbitt}}]{Cullen_subSQL}%
  \BibitemOpen
  \bibfield  {author} {\bibinfo {author} {\bibfnamefont {T.}~\bibnamefont {Cullen}}, \bibinfo {author} {\bibfnamefont {R.}~\bibnamefont {Pagano}}, \bibinfo {author} {\bibfnamefont {S.}~\bibnamefont {Aronson}}, \bibinfo {author} {\bibfnamefont {J.}~\bibnamefont {Cripe}}, \bibinfo {author} {\bibfnamefont {S.~S.}\ \bibnamefont {Sharif}}, \bibinfo {author} {\bibfnamefont {M.}~\bibnamefont {Lollie}}, \bibinfo {author} {\bibfnamefont {H.}~\bibnamefont {Cain}}, \bibinfo {author} {\bibfnamefont {P.}~\bibnamefont {Heu}}, \bibinfo {author} {\bibfnamefont {D.}~\bibnamefont {Follman}}, \bibinfo {author} {\bibfnamefont {G.~D.}\ \bibnamefont {Cole}}, \bibinfo {author} {\bibfnamefont {N.}~\bibnamefont {Aggarwal}},\ and\ \bibinfo {author} {\bibfnamefont {T.}~\bibnamefont {Corbitt}},\ }\bibfield  {journal} {\bibinfo  {journal} {Physical Review Letters}\ }\textbf {\bibinfo {volume} {133}},\ \href {https://doi.org/10.1103/PhysRevLett.133.113602} {10.1103/PhysRevLett.133.113602} (\bibinfo {year} {2024})\BibitemShut {NoStop}%
\bibitem [{\citenamefont {Unruh}(1983{\natexlab{a}})}]{Unruh_NATO1}%
  \BibitemOpen
  \bibfield  {author} {\bibinfo {author} {\bibfnamefont {W.~G.}\ \bibnamefont {Unruh}},\ }\bibinfo {title} {Readout state preparation and quantum non-demolition},\ in\ \href {https://doi.org/10.1007/978-1-4613-3712-6_27} {\emph {\bibinfo {booktitle} {Quantum Optics, Experimental Gravity, and Measurement Theory}}},\ \bibinfo {editor} {edited by\ \bibinfo {editor} {\bibfnamefont {P.}~\bibnamefont {Meystre}}\ and\ \bibinfo {editor} {\bibfnamefont {M.~O.}\ \bibnamefont {Scully}}}\ (\bibinfo  {publisher} {Springer US},\ \bibinfo {address} {Boston, MA},\ \bibinfo {year} {1983})\ pp.\ \bibinfo {pages} {637--645}\BibitemShut {NoStop}%
\bibitem [{\citenamefont {Unruh}(1983{\natexlab{b}})}]{Unruh_NATO2}%
  \BibitemOpen
  \bibfield  {author} {\bibinfo {author} {\bibfnamefont {W.~G.}\ \bibnamefont {Unruh}},\ }\bibinfo {title} {Quantum noise in the interferometer detector},\ in\ \href {https://doi.org/10.1007/978-1-4613-3712-6_28} {\emph {\bibinfo {booktitle} {Quantum Optics, Experimental Gravity, and Measurement Theory}}},\ \bibinfo {editor} {edited by\ \bibinfo {editor} {\bibfnamefont {P.}~\bibnamefont {Meystre}}\ and\ \bibinfo {editor} {\bibfnamefont {M.~O.}\ \bibnamefont {Scully}}}\ (\bibinfo  {publisher} {Springer US},\ \bibinfo {address} {Boston, MA},\ \bibinfo {year} {1983})\ pp.\ \bibinfo {pages} {647--660}\BibitemShut {NoStop}%
\bibitem [{\citenamefont {Kimble}\ \emph {et~al.}(2001)\citenamefont {Kimble}, \citenamefont {Levin}, \citenamefont {Matsko}, \citenamefont {Thorne},\ and\ \citenamefont {Vyatchanin}}]{Kimble}%
  \BibitemOpen
  \bibfield  {author} {\bibinfo {author} {\bibfnamefont {H.~J.}\ \bibnamefont {Kimble}}, \bibinfo {author} {\bibfnamefont {Y.}~\bibnamefont {Levin}}, \bibinfo {author} {\bibfnamefont {A.~B.}\ \bibnamefont {Matsko}}, \bibinfo {author} {\bibfnamefont {K.~S.}\ \bibnamefont {Thorne}},\ and\ \bibinfo {author} {\bibfnamefont {S.~P.}\ \bibnamefont {Vyatchanin}},\ }\href {https://doi.org/10.1103/PhysRevD.65.022002} {\bibfield  {journal} {\bibinfo  {journal} {Physical Review D}\ }\textbf {\bibinfo {volume} {65}},\ \bibinfo {pages} {022002} (\bibinfo {year} {2001})}\BibitemShut {NoStop}%
\bibitem [{\citenamefont {Mason}\ \emph {et~al.}(2019)\citenamefont {Mason}, \citenamefont {Chen}, \citenamefont {Rossi}, \citenamefont {Tsaturyan},\ and\ \citenamefont {Schliesser}}]{Mason_subSQL}%
  \BibitemOpen
  \bibfield  {author} {\bibinfo {author} {\bibfnamefont {D.}~\bibnamefont {Mason}}, \bibinfo {author} {\bibfnamefont {J.}~\bibnamefont {Chen}}, \bibinfo {author} {\bibfnamefont {M.}~\bibnamefont {Rossi}}, \bibinfo {author} {\bibfnamefont {Y.}~\bibnamefont {Tsaturyan}},\ and\ \bibinfo {author} {\bibfnamefont {A.}~\bibnamefont {Schliesser}},\ }\href {https://doi.org/10.1038/s41567-019-0533-5} {\bibfield  {journal} {\bibinfo  {journal} {Nature Physics}\ }\textbf {\bibinfo {volume} {15}},\ \bibinfo {pages} {745} (\bibinfo {year} {2019})}\BibitemShut {NoStop}%
\bibitem [{\citenamefont {Yu}\ \emph {et~al.}(2020)\citenamefont {Yu}, \citenamefont {McCuller}, \citenamefont {Tse},\ and\ \citenamefont {et~al}}]{Yu_subSQL}%
  \BibitemOpen
  \bibfield  {author} {\bibinfo {author} {\bibfnamefont {H.}~\bibnamefont {Yu}}, \bibinfo {author} {\bibfnamefont {L.}~\bibnamefont {McCuller}}, \bibinfo {author} {\bibfnamefont {M.}~\bibnamefont {Tse}},\ and\ \bibinfo {author} {\bibnamefont {et~al}},\ }\href {https://doi.org/10.1038/s41586-020-2420-8} {\bibfield  {journal} {\bibinfo  {journal} {Nature}\ }\textbf {\bibinfo {volume} {583}},\ \bibinfo {pages} {43} (\bibinfo {year} {2020})}\BibitemShut {NoStop}%
\bibitem [{\citenamefont {Jia}\ and\ \citenamefont {et. al.}(2024)}]{jia_conditional_subsql}%
  \BibitemOpen
  \bibfield  {author} {\bibinfo {author} {\bibfnamefont {W.}~\bibnamefont {Jia}}\ and\ \bibinfo {author} {\bibnamefont {et. al.}},\ }\href {https://doi.org/10.1126/science.ado8069} {\bibfield  {journal} {\bibinfo  {journal} {Science}\ }\textbf {\bibinfo {volume} {385}},\ \bibinfo {pages} {1318–1321} (\bibinfo {year} {2024})}\BibitemShut {NoStop}%
\bibitem [{\citenamefont {Gras}\ \emph {et~al.}(2017)\citenamefont {Gras}, \citenamefont {Yu}, \citenamefont {Yam}, \citenamefont {Martynov},\ and\ \citenamefont {Evans}}]{LIGO_thermal_noise_msr1}%
  \BibitemOpen
  \bibfield  {author} {\bibinfo {author} {\bibfnamefont {S.}~\bibnamefont {Gras}}, \bibinfo {author} {\bibfnamefont {H.}~\bibnamefont {Yu}}, \bibinfo {author} {\bibfnamefont {W.}~\bibnamefont {Yam}}, \bibinfo {author} {\bibfnamefont {D.}~\bibnamefont {Martynov}},\ and\ \bibinfo {author} {\bibfnamefont {M.}~\bibnamefont {Evans}},\ }\bibfield  {journal} {\bibinfo  {journal} {Physical Review D}\ }\textbf {\bibinfo {volume} {95}},\ \href {https://doi.org/10.1103/PhysRevD.95.022001} {10.1103/PhysRevD.95.022001} (\bibinfo {year} {2017})\BibitemShut {NoStop}%
\bibitem [{\citenamefont {Gras}\ and\ \citenamefont {Evans}(2018)}]{LIGO_thermal_noise_msr2}%
  \BibitemOpen
  \bibfield  {author} {\bibinfo {author} {\bibfnamefont {S.}~\bibnamefont {Gras}}\ and\ \bibinfo {author} {\bibfnamefont {M.}~\bibnamefont {Evans}},\ }\bibfield  {journal} {\bibinfo  {journal} {Physical Review D}\ }\textbf {\bibinfo {volume} {98}},\ \href {https://doi.org/10.1103/PhysRevD.98.122001} {10.1103/PhysRevD.98.122001} (\bibinfo {year} {2018})\BibitemShut {NoStop}%
\bibitem [{\citenamefont {Cripe}(2018)}]{thesis_cripe}%
  \BibitemOpen
  \bibfield  {author} {\bibinfo {author} {\bibfnamefont {J.~D.}\ \bibnamefont {Cripe}},\ }\emph {\bibinfo {title} {Broadband Measurement and Reduction of Quantum Radiation Pressure Noise in the Audio Band}},\ \href@noop {} {Ph.D. thesis},\ \bibinfo  {school} {Louisiana State University} (\bibinfo {year} {2018})\BibitemShut {NoStop}%
\bibitem [{\citenamefont {Miyakawa}\ \emph {et~al.}(2006)\citenamefont {Miyakawa}, \citenamefont {Ward}, \citenamefont {Adhikari}, \citenamefont {Evans}, \citenamefont {Abbott}, \citenamefont {Bork}, \citenamefont {Busby}, \citenamefont {Heefner}, \citenamefont {Ivanov}, \citenamefont {Smith}, \citenamefont {Taylor}, \citenamefont {Vass}, \citenamefont {Weinstein}, \citenamefont {Varvella}, \citenamefont {Kawamura}, \citenamefont {Kawazoe}, \citenamefont {Sakata},\ and\ \citenamefont {Mow-Lowry}}]{Miyakawa_40m_os}%
  \BibitemOpen
  \bibfield  {author} {\bibinfo {author} {\bibfnamefont {O.}~\bibnamefont {Miyakawa}}, \bibinfo {author} {\bibfnamefont {R.}~\bibnamefont {Ward}}, \bibinfo {author} {\bibfnamefont {R.}~\bibnamefont {Adhikari}}, \bibinfo {author} {\bibfnamefont {M.}~\bibnamefont {Evans}}, \bibinfo {author} {\bibfnamefont {B.}~\bibnamefont {Abbott}}, \bibinfo {author} {\bibfnamefont {R.}~\bibnamefont {Bork}}, \bibinfo {author} {\bibfnamefont {D.}~\bibnamefont {Busby}}, \bibinfo {author} {\bibfnamefont {J.}~\bibnamefont {Heefner}}, \bibinfo {author} {\bibfnamefont {A.}~\bibnamefont {Ivanov}}, \bibinfo {author} {\bibfnamefont {M.}~\bibnamefont {Smith}}, \bibinfo {author} {\bibfnamefont {R.}~\bibnamefont {Taylor}}, \bibinfo {author} {\bibfnamefont {S.}~\bibnamefont {Vass}}, \bibinfo {author} {\bibfnamefont {A.}~\bibnamefont {Weinstein}}, \bibinfo {author} {\bibfnamefont {M.}~\bibnamefont {Varvella}}, \bibinfo {author} {\bibfnamefont {S.}~\bibnamefont {Kawamura}}, \bibinfo {author} {\bibfnamefont {F.}~\bibnamefont {Kawazoe}},
  \bibinfo {author} {\bibfnamefont {S.}~\bibnamefont {Sakata}},\ and\ \bibinfo {author} {\bibfnamefont {C.}~\bibnamefont {Mow-Lowry}},\ }\href {https://doi.org/10.1103/PhysRevD.74.022001} {\bibfield  {journal} {\bibinfo  {journal} {Physical Review D}\ }\textbf {\bibinfo {volume} {74}},\ \bibinfo {pages} {022001} (\bibinfo {year} {2006})}\BibitemShut {NoStop}%
\bibitem [{\citenamefont {Buonanno}\ and\ \citenamefont {Chen}(2002)}]{Chen_ligo_optical_spring}%
  \BibitemOpen
  \bibfield  {author} {\bibinfo {author} {\bibfnamefont {A.}~\bibnamefont {Buonanno}}\ and\ \bibinfo {author} {\bibfnamefont {Y.}~\bibnamefont {Chen}},\ }\href {http://stacks.iop.org/0264-9381/19/i=7/a=346} {\bibfield  {journal} {\bibinfo  {journal} {Classical and Quantum Gravity}\ }\textbf {\bibinfo {volume} {19}},\ \bibinfo {pages} {1569} (\bibinfo {year} {2002})}\BibitemShut {NoStop}%
\bibitem [{\citenamefont {Tso}\ \emph {et~al.}(2019)\citenamefont {Tso}, \citenamefont {Gerosa},\ and\ \citenamefont {Chen}}]{Chen_lisa_forewarnings}%
  \BibitemOpen
  \bibfield  {author} {\bibinfo {author} {\bibfnamefont {R.}~\bibnamefont {Tso}}, \bibinfo {author} {\bibfnamefont {D.}~\bibnamefont {Gerosa}},\ and\ \bibinfo {author} {\bibfnamefont {Y.}~\bibnamefont {Chen}},\ }\bibfield  {journal} {\bibinfo  {journal} {Physical Review D}\ }\textbf {\bibinfo {volume} {99}},\ \href {https://doi.org/10.1103/PhysRevD.99.124043} {10.1103/PhysRevD.99.124043} (\bibinfo {year} {2019})\BibitemShut {NoStop}%
\bibitem [{\citenamefont {Aronson}\ \emph {et~al.}(2024)\citenamefont {Aronson}, \citenamefont {Pagano}, \citenamefont {Cullen}, \citenamefont {Cole},\ and\ \citenamefont {Corbitt}}]{Aronson_op_spring_tracking}%
  \BibitemOpen
  \bibfield  {author} {\bibinfo {author} {\bibfnamefont {S.}~\bibnamefont {Aronson}}, \bibinfo {author} {\bibfnamefont {R.}~\bibnamefont {Pagano}}, \bibinfo {author} {\bibfnamefont {T.}~\bibnamefont {Cullen}}, \bibinfo {author} {\bibfnamefont {G.~D.}\ \bibnamefont {Cole}},\ and\ \bibinfo {author} {\bibfnamefont {T.}~\bibnamefont {Corbitt}},\ }\href {https://doi.org/10.1364/OL.540195} {\bibfield  {journal} {\bibinfo  {journal} {Optics Letters}\ }\textbf {\bibinfo {volume} {49}},\ \bibinfo {pages} {6980} (\bibinfo {year} {2024})}\BibitemShut {NoStop}%
\bibitem [{\citenamefont {Carney}\ \emph {et~al.}(2021)\citenamefont {Carney}, \citenamefont {Krnjaic}, \citenamefont {Moore}, \citenamefont {Regal}, \citenamefont {Afek}, \citenamefont {Bhave}, \citenamefont {Brubaker}, \citenamefont {Corbitt}, \citenamefont {Cripe}, \citenamefont {Crisosto}, \citenamefont {Geraci}, \citenamefont {Ghosh}, \citenamefont {Harris}, \citenamefont {Hook}, \citenamefont {Kolb}, \citenamefont {Kunjummen}, \citenamefont {Lang}, \citenamefont {Li}, \citenamefont {Lin}, \citenamefont {Liu}, \citenamefont {Lykken}, \citenamefont {Magrini}, \citenamefont {Manley}, \citenamefont {Matsumoto}, \citenamefont {Monte}, \citenamefont {Monteiro}, \citenamefont {Purdy}, \citenamefont {Riedel}, \citenamefont {Singh}, \citenamefont {Singh}, \citenamefont {Sinha}, \citenamefont {Taylor}, \citenamefont {Qin}, \citenamefont {Wilson},\ and\ \citenamefont {Zhao}}]{Carney_mechanical_dark_matter_sensor}%
  \BibitemOpen
  \bibfield  {author} {\bibinfo {author} {\bibfnamefont {D.}~\bibnamefont {Carney}}, \bibinfo {author} {\bibfnamefont {G.}~\bibnamefont {Krnjaic}}, \bibinfo {author} {\bibfnamefont {D.~C.}\ \bibnamefont {Moore}}, \bibinfo {author} {\bibfnamefont {C.~A.}\ \bibnamefont {Regal}}, \bibinfo {author} {\bibfnamefont {G.}~\bibnamefont {Afek}}, \bibinfo {author} {\bibfnamefont {S.}~\bibnamefont {Bhave}}, \bibinfo {author} {\bibfnamefont {B.}~\bibnamefont {Brubaker}}, \bibinfo {author} {\bibfnamefont {T.}~\bibnamefont {Corbitt}}, \bibinfo {author} {\bibfnamefont {J.}~\bibnamefont {Cripe}}, \bibinfo {author} {\bibfnamefont {N.}~\bibnamefont {Crisosto}}, \bibinfo {author} {\bibfnamefont {A.}~\bibnamefont {Geraci}}, \bibinfo {author} {\bibfnamefont {S.}~\bibnamefont {Ghosh}}, \bibinfo {author} {\bibfnamefont {J.~G.~E.}\ \bibnamefont {Harris}}, \bibinfo {author} {\bibfnamefont {A.}~\bibnamefont {Hook}}, \bibinfo {author} {\bibfnamefont {E.~W.}\ \bibnamefont {Kolb}}, \bibinfo {author} {\bibfnamefont {J.}~\bibnamefont
  {Kunjummen}}, \bibinfo {author} {\bibfnamefont {R.~F.}\ \bibnamefont {Lang}}, \bibinfo {author} {\bibfnamefont {T.}~\bibnamefont {Li}}, \bibinfo {author} {\bibfnamefont {T.}~\bibnamefont {Lin}}, \bibinfo {author} {\bibfnamefont {Z.}~\bibnamefont {Liu}}, \bibinfo {author} {\bibfnamefont {J.}~\bibnamefont {Lykken}}, \bibinfo {author} {\bibfnamefont {L.}~\bibnamefont {Magrini}}, \bibinfo {author} {\bibfnamefont {J.}~\bibnamefont {Manley}}, \bibinfo {author} {\bibfnamefont {N.}~\bibnamefont {Matsumoto}}, \bibinfo {author} {\bibfnamefont {A.}~\bibnamefont {Monte}}, \bibinfo {author} {\bibfnamefont {F.}~\bibnamefont {Monteiro}}, \bibinfo {author} {\bibfnamefont {T.}~\bibnamefont {Purdy}}, \bibinfo {author} {\bibfnamefont {C.~J.}\ \bibnamefont {Riedel}}, \bibinfo {author} {\bibfnamefont {R.}~\bibnamefont {Singh}}, \bibinfo {author} {\bibfnamefont {S.}~\bibnamefont {Singh}}, \bibinfo {author} {\bibfnamefont {K.}~\bibnamefont {Sinha}}, \bibinfo {author} {\bibfnamefont {J.~M.}\ \bibnamefont {Taylor}}, \bibinfo
  {author} {\bibfnamefont {J.}~\bibnamefont {Qin}}, \bibinfo {author} {\bibfnamefont {D.~J.}\ \bibnamefont {Wilson}},\ and\ \bibinfo {author} {\bibfnamefont {Y.}~\bibnamefont {Zhao}},\ }\href {https://doi.org/10.1088/2058-9565/abcfcd} {\bibfield  {journal} {\bibinfo  {journal} {Quantum Science and Technology}\ }\textbf {\bibinfo {volume} {6}},\ \bibinfo {pages} {024002} (\bibinfo {year} {2021})}\BibitemShut {NoStop}%
\bibitem [{\citenamefont {Collaboration}\ \emph {et~al.}(2022)\citenamefont {Collaboration}, \citenamefont {Attanasio}, \citenamefont {Bhave}, \citenamefont {Blanco}, \citenamefont {Carney}, \citenamefont {Demarteau}, \citenamefont {Elshimy}, \citenamefont {Febbraro}, \citenamefont {Feldman}, \citenamefont {Ghosh}, \citenamefont {Hickin}, \citenamefont {Hong}, \citenamefont {Lang}, \citenamefont {Lawrie}, \citenamefont {Li}, \citenamefont {Liu}, \citenamefont {Maldonado}, \citenamefont {Marvinney}, \citenamefont {Oo}, \citenamefont {Pai}, \citenamefont {Pooser}, \citenamefont {Qin}, \citenamefont {Sparmann}, \citenamefont {Taylor}, \citenamefont {Tian},\ and\ \citenamefont {Tunnell}}]{snowmass2021whitepaper}%
  \BibitemOpen
  \bibfield  {author} {\bibinfo {author} {\bibfnamefont {T.~W.}\ \bibnamefont {Collaboration}}, \bibinfo {author} {\bibfnamefont {A.}~\bibnamefont {Attanasio}}, \bibinfo {author} {\bibfnamefont {S.~A.}\ \bibnamefont {Bhave}}, \bibinfo {author} {\bibfnamefont {C.}~\bibnamefont {Blanco}}, \bibinfo {author} {\bibfnamefont {D.}~\bibnamefont {Carney}}, \bibinfo {author} {\bibfnamefont {M.}~\bibnamefont {Demarteau}}, \bibinfo {author} {\bibfnamefont {B.}~\bibnamefont {Elshimy}}, \bibinfo {author} {\bibfnamefont {M.}~\bibnamefont {Febbraro}}, \bibinfo {author} {\bibfnamefont {M.~A.}\ \bibnamefont {Feldman}}, \bibinfo {author} {\bibfnamefont {S.}~\bibnamefont {Ghosh}}, \bibinfo {author} {\bibfnamefont {A.}~\bibnamefont {Hickin}}, \bibinfo {author} {\bibfnamefont {S.}~\bibnamefont {Hong}}, \bibinfo {author} {\bibfnamefont {R.~F.}\ \bibnamefont {Lang}}, \bibinfo {author} {\bibfnamefont {B.}~\bibnamefont {Lawrie}}, \bibinfo {author} {\bibfnamefont {S.}~\bibnamefont {Li}}, \bibinfo {author} {\bibfnamefont
  {Z.}~\bibnamefont {Liu}}, \bibinfo {author} {\bibfnamefont {J.~P.~A.}\ \bibnamefont {Maldonado}}, \bibinfo {author} {\bibfnamefont {C.}~\bibnamefont {Marvinney}}, \bibinfo {author} {\bibfnamefont {H.~Z.~Y.}\ \bibnamefont {Oo}}, \bibinfo {author} {\bibfnamefont {Y.-Y.}\ \bibnamefont {Pai}}, \bibinfo {author} {\bibfnamefont {R.}~\bibnamefont {Pooser}}, \bibinfo {author} {\bibfnamefont {J.}~\bibnamefont {Qin}}, \bibinfo {author} {\bibfnamefont {T.~J.}\ \bibnamefont {Sparmann}}, \bibinfo {author} {\bibfnamefont {J.~M.}\ \bibnamefont {Taylor}}, \bibinfo {author} {\bibfnamefont {H.}~\bibnamefont {Tian}},\ and\ \bibinfo {author} {\bibfnamefont {C.}~\bibnamefont {Tunnell}},\ }\href {https://arxiv.org/abs/2203.07242} {\bibinfo {title} {Snowmass 2021 white paper: The windchime project}} (\bibinfo {year} {2022}),\ \Eprint {https://arxiv.org/abs/2203.07242} {arXiv:2203.07242 [hep-ex]} \BibitemShut {NoStop}%
\bibitem [{\citenamefont {Cole}\ \emph {et~al.}(2013)\citenamefont {Cole}, \citenamefont {Zhang}, \citenamefont {Martin}, \citenamefont {Ye},\ and\ \citenamefont {Aspelmeyer}}]{Cole_10fold_reduction}%
  \BibitemOpen
  \bibfield  {author} {\bibinfo {author} {\bibfnamefont {G.~D.}\ \bibnamefont {Cole}}, \bibinfo {author} {\bibfnamefont {W.}~\bibnamefont {Zhang}}, \bibinfo {author} {\bibfnamefont {M.~J.}\ \bibnamefont {Martin}}, \bibinfo {author} {\bibfnamefont {J.}~\bibnamefont {Ye}},\ and\ \bibinfo {author} {\bibfnamefont {M.}~\bibnamefont {Aspelmeyer}},\ }\href {https://doi.org/10.1038/NPHOTON.2013.174} {\bibfield  {journal} {\bibinfo  {journal} {Nature Photonics}\ }\textbf {\bibinfo {volume} {7}},\ \bibinfo {pages} {644} (\bibinfo {year} {2013})}\BibitemShut {NoStop}%
\bibitem [{\citenamefont {Cole}\ \emph {et~al.}(2023)\citenamefont {Cole}, \citenamefont {Ballmer}, \citenamefont {Billingsley}, \citenamefont {Cataño-Lopez}, \citenamefont {Fejer}, \citenamefont {Fritschel}, \citenamefont {Gretarsson}, \citenamefont {Harry}, \citenamefont {Kedar}, \citenamefont {Legero}, \citenamefont {Makarem}, \citenamefont {Penn}, \citenamefont {Reitze}, \citenamefont {Steinlechner}, \citenamefont {Sterr}, \citenamefont {Tanioka}, \citenamefont {Truong}, \citenamefont {Ye},\ and\ \citenamefont {Yu}}]{coal2023_algaas_in_gw_detectors}%
  \BibitemOpen
  \bibfield  {author} {\bibinfo {author} {\bibfnamefont {G.~D.}\ \bibnamefont {Cole}}, \bibinfo {author} {\bibfnamefont {S.~W.}\ \bibnamefont {Ballmer}}, \bibinfo {author} {\bibfnamefont {G.}~\bibnamefont {Billingsley}}, \bibinfo {author} {\bibfnamefont {S.~B.}\ \bibnamefont {Cataño-Lopez}}, \bibinfo {author} {\bibfnamefont {M.}~\bibnamefont {Fejer}}, \bibinfo {author} {\bibfnamefont {P.}~\bibnamefont {Fritschel}}, \bibinfo {author} {\bibfnamefont {A.~M.}\ \bibnamefont {Gretarsson}}, \bibinfo {author} {\bibfnamefont {G.~M.}\ \bibnamefont {Harry}}, \bibinfo {author} {\bibfnamefont {D.}~\bibnamefont {Kedar}}, \bibinfo {author} {\bibfnamefont {T.}~\bibnamefont {Legero}}, \bibinfo {author} {\bibfnamefont {C.}~\bibnamefont {Makarem}}, \bibinfo {author} {\bibfnamefont {S.~D.}\ \bibnamefont {Penn}}, \bibinfo {author} {\bibfnamefont {D.~H.}\ \bibnamefont {Reitze}}, \bibinfo {author} {\bibfnamefont {J.}~\bibnamefont {Steinlechner}}, \bibinfo {author} {\bibfnamefont {U.}~\bibnamefont {Sterr}}, \bibinfo {author}
  {\bibfnamefont {S.}~\bibnamefont {Tanioka}}, \bibinfo {author} {\bibfnamefont {G.-W.}\ \bibnamefont {Truong}}, \bibinfo {author} {\bibfnamefont {J.}~\bibnamefont {Ye}},\ and\ \bibinfo {author} {\bibfnamefont {J.}~\bibnamefont {Yu}},\ }\href {https://doi.org/10.1063/5.0140663} {\bibfield  {journal} {\bibinfo  {journal} {Applied Physics Letters}\ }\textbf {\bibinfo {volume} {122}},\ \bibinfo {pages} {110502} (\bibinfo {year} {2023})},\ \Eprint {https://arxiv.org/abs/https://pubs.aip.org/aip/apl/article-pdf/doi/10.1063/5.0140663/19999213/110502\_1\_5.0140663.pdf} {https://pubs.aip.org/aip/apl/article-pdf/doi/10.1063/5.0140663/19999213/110502\_1\_5.0140663.pdf} \BibitemShut {NoStop}%
\bibitem [{\citenamefont {Aggarwal}\ \emph {et~al.}(2020)\citenamefont {Aggarwal}, \citenamefont {Cullen}, \citenamefont {Cripe},\ and\ \citenamefont {\textit{et al}.}}]{nancy_squeeze}%
  \BibitemOpen
  \bibfield  {author} {\bibinfo {author} {\bibfnamefont {N.}~\bibnamefont {Aggarwal}}, \bibinfo {author} {\bibfnamefont {T.}~\bibnamefont {Cullen}}, \bibinfo {author} {\bibfnamefont {J.}~\bibnamefont {Cripe}},\ and\ \bibinfo {author} {\bibnamefont {\textit{et al}.}},\ }\href {https://doi.org/https://doi.org/10.1038/s41567-020-0877-x} {\bibfield  {journal} {\bibinfo  {journal} {Nature Physics}\ }\textbf {\bibinfo {volume} {16}},\ \bibinfo {pages} {784–788} (\bibinfo {year} {2020})}\BibitemShut {NoStop}%
\bibitem [{\citenamefont {Cripe}\ \emph {et~al.}(2019)\citenamefont {Cripe}, \citenamefont {Aggarwal}, \citenamefont {Lanza}, \citenamefont {Libson}, \citenamefont {Singh}, \citenamefont {Heu}, \citenamefont {Follman}, \citenamefont {Cole}, \citenamefont {Mavalvala},\ and\ \citenamefont {Corbitt}}]{Cripe_QRPN}%
  \BibitemOpen
  \bibfield  {author} {\bibinfo {author} {\bibfnamefont {J.}~\bibnamefont {Cripe}}, \bibinfo {author} {\bibfnamefont {N.}~\bibnamefont {Aggarwal}}, \bibinfo {author} {\bibfnamefont {R.}~\bibnamefont {Lanza}}, \bibinfo {author} {\bibfnamefont {A.}~\bibnamefont {Libson}}, \bibinfo {author} {\bibfnamefont {R.}~\bibnamefont {Singh}}, \bibinfo {author} {\bibfnamefont {P.}~\bibnamefont {Heu}}, \bibinfo {author} {\bibfnamefont {D.}~\bibnamefont {Follman}}, \bibinfo {author} {\bibfnamefont {G.~D.}\ \bibnamefont {Cole}}, \bibinfo {author} {\bibfnamefont {N.}~\bibnamefont {Mavalvala}},\ and\ \bibinfo {author} {\bibfnamefont {T.}~\bibnamefont {Corbitt}},\ }\href {https://doi.org/10.1038/s41586-019-1051-4} {\bibfield  {journal} {\bibinfo  {journal} {Nature}\ }\textbf {\bibinfo {volume} {568}},\ \bibinfo {pages} {364} (\bibinfo {year} {2019})}\BibitemShut {NoStop}%
\bibitem [{\citenamefont {Yap}\ \emph {et~al.}(2020)\citenamefont {Yap}, \citenamefont {Cripe}, \citenamefont {Mansell}, \citenamefont {McRae}, \citenamefont {Ward}, \citenamefont {Slagmolen}, \citenamefont {Heu}, \citenamefont {Follman}, \citenamefont {Cole}, \citenamefont {Corbitt},\ and\ \citenamefont {McClelland}}]{Yap_qrpn_reduction_w_squeezed_light}%
  \BibitemOpen
  \bibfield  {author} {\bibinfo {author} {\bibfnamefont {M.~J.}\ \bibnamefont {Yap}}, \bibinfo {author} {\bibfnamefont {J.}~\bibnamefont {Cripe}}, \bibinfo {author} {\bibfnamefont {G.~L.}\ \bibnamefont {Mansell}}, \bibinfo {author} {\bibfnamefont {T.~G.}\ \bibnamefont {McRae}}, \bibinfo {author} {\bibfnamefont {R.~L.}\ \bibnamefont {Ward}}, \bibinfo {author} {\bibfnamefont {B.~J.~J.}\ \bibnamefont {Slagmolen}}, \bibinfo {author} {\bibfnamefont {P.}~\bibnamefont {Heu}}, \bibinfo {author} {\bibfnamefont {D.}~\bibnamefont {Follman}}, \bibinfo {author} {\bibfnamefont {G.~D.}\ \bibnamefont {Cole}}, \bibinfo {author} {\bibfnamefont {T.}~\bibnamefont {Corbitt}},\ and\ \bibinfo {author} {\bibfnamefont {D.~E.}\ \bibnamefont {McClelland}},\ }\href {https://doi.org/10.1038/s41566-019-0527-y} {\bibfield  {journal} {\bibinfo  {journal} {Nature Photonics}\ }\textbf {\bibinfo {volume} {14}},\ \bibinfo {pages} {19} (\bibinfo {year} {2020})}\BibitemShut {NoStop}%
\bibitem [{\citenamefont {Cripe}\ \emph {et~al.}(2020)\citenamefont {Cripe}, \citenamefont {Cullen}, \citenamefont {Chen}, \citenamefont {Heu}, \citenamefont {Follman}, \citenamefont {Cole},\ and\ \citenamefont {Corbitt}}]{Cripe_QBAcancellation}%
  \BibitemOpen
  \bibfield  {author} {\bibinfo {author} {\bibfnamefont {J.}~\bibnamefont {Cripe}}, \bibinfo {author} {\bibfnamefont {T.}~\bibnamefont {Cullen}}, \bibinfo {author} {\bibfnamefont {Y.}~\bibnamefont {Chen}}, \bibinfo {author} {\bibfnamefont {P.}~\bibnamefont {Heu}}, \bibinfo {author} {\bibfnamefont {D.}~\bibnamefont {Follman}}, \bibinfo {author} {\bibfnamefont {G.~D.}\ \bibnamefont {Cole}},\ and\ \bibinfo {author} {\bibfnamefont {T.}~\bibnamefont {Corbitt}},\ }\href {https://doi.org/10.1103/PhysRevX.10.031065} {\bibfield  {journal} {\bibinfo  {journal} {Physical Review X}\ }\textbf {\bibinfo {volume} {10}},\ \bibinfo {pages} {031065} (\bibinfo {year} {2020})}\BibitemShut {NoStop}%
\bibitem [{\citenamefont {Saulson}(1990)}]{Saulson_thermal_noise}%
  \BibitemOpen
  \bibfield  {author} {\bibinfo {author} {\bibfnamefont {P.~R.}\ \bibnamefont {Saulson}},\ }\href {https://doi.org/10.1103/PhysRevD.42.2437} {\bibfield  {journal} {\bibinfo  {journal} {Physical Review D}\ }\textbf {\bibinfo {volume} {42}},\ \bibinfo {pages} {2437} (\bibinfo {year} {1990})}\BibitemShut {NoStop}%
\bibitem [{\citenamefont {Gabriela}\ and\ \citenamefont {Saulson}(1994)}]{Gonzalez_thermal_noise}%
  \BibitemOpen
  \bibfield  {author} {\bibinfo {author} {\bibfnamefont {G.}~\bibnamefont {Gabriela}}\ and\ \bibinfo {author} {\bibfnamefont {P.}~\bibnamefont {Saulson}},\ }\href {https://doi.org/10.1121/1.410467} {\bibfield  {journal} {\bibinfo  {journal} {Journal Of The Acoustical Society Of America}\ }\textbf {\bibinfo {volume} {96}},\ \bibinfo {pages} {207} (\bibinfo {year} {1994})}\BibitemShut {NoStop}%
\bibitem [{\citenamefont {Saulson}(1994)}]{Saulson_fundamentals}%
  \BibitemOpen
  \bibfield  {author} {\bibinfo {author} {\bibfnamefont {P.~R.}\ \bibnamefont {Saulson}},\ }\href@noop {} {\emph {\bibinfo {title} {Fundamentals of Interferometric Gravitational Wave Detectors"}}}\ (\bibinfo  {publisher} {World Scientific},\ \bibinfo {address} {River Edge, N.J},\ \bibinfo {year} {1994})\BibitemShut {NoStop}%
\bibitem [{\citenamefont {Martynov}\ and\ \citenamefont {et. al.}(2017)}]{Martynov_cpsd_measurement}%
  \BibitemOpen
  \bibfield  {author} {\bibinfo {author} {\bibfnamefont {D.~V.}\ \bibnamefont {Martynov}}\ and\ \bibinfo {author} {\bibnamefont {et. al.}},\ }\bibfield  {journal} {\bibinfo  {journal} {Physical Review A}\ }\textbf {\bibinfo {volume} {95}},\ \href {https://doi.org/10.1103/PhysRevA.95.043831} {10.1103/PhysRevA.95.043831} (\bibinfo {year} {2017})\BibitemShut {NoStop}%
\bibitem [{\citenamefont {Tanioka}\ \emph {et~al.}(2023)\citenamefont {Tanioka}, \citenamefont {Vander-Hyde}, \citenamefont {Cole}, \citenamefont {Penn},\ and\ \citenamefont {Ballmer}}]{Tanioka_electro_optic_AlGaAs}%
  \BibitemOpen
  \bibfield  {author} {\bibinfo {author} {\bibfnamefont {S.}~\bibnamefont {Tanioka}}, \bibinfo {author} {\bibfnamefont {D.}~\bibnamefont {Vander-Hyde}}, \bibinfo {author} {\bibfnamefont {G.~D.}\ \bibnamefont {Cole}}, \bibinfo {author} {\bibfnamefont {S.~D.}\ \bibnamefont {Penn}},\ and\ \bibinfo {author} {\bibfnamefont {S.~W.}\ \bibnamefont {Ballmer}},\ }\bibfield  {journal} {\bibinfo  {journal} {Physical Review D}\ }\textbf {\bibinfo {volume} {107}},\ \href {https://doi.org/10.1103/PhysRevD.107.022003} {10.1103/PhysRevD.107.022003} (\bibinfo {year} {2023})\BibitemShut {NoStop}%
\bibitem [{\citenamefont {Yu}\ \emph {et~al.}(2023)\citenamefont {Yu}, \citenamefont {Hafner}, \citenamefont {Legero}, \citenamefont {Herbers}, \citenamefont {Nicolodi}, \citenamefont {Ma}, \citenamefont {Riehle}, \citenamefont {Sterr}, \citenamefont {Kedar}, \citenamefont {Robinson}, \citenamefont {Oelker},\ and\ \citenamefont {Ye}}]{Yu_birefringent_noise}%
  \BibitemOpen
  \bibfield  {author} {\bibinfo {author} {\bibfnamefont {J.}~\bibnamefont {Yu}}, \bibinfo {author} {\bibfnamefont {S.}~\bibnamefont {Hafner}}, \bibinfo {author} {\bibfnamefont {T.}~\bibnamefont {Legero}}, \bibinfo {author} {\bibfnamefont {S.}~\bibnamefont {Herbers}}, \bibinfo {author} {\bibfnamefont {D.}~\bibnamefont {Nicolodi}}, \bibinfo {author} {\bibfnamefont {C.~Y.}\ \bibnamefont {Ma}}, \bibinfo {author} {\bibfnamefont {F.}~\bibnamefont {Riehle}}, \bibinfo {author} {\bibfnamefont {U.}~\bibnamefont {Sterr}}, \bibinfo {author} {\bibfnamefont {D.}~\bibnamefont {Kedar}}, \bibinfo {author} {\bibfnamefont {J.~M.}\ \bibnamefont {Robinson}}, \bibinfo {author} {\bibfnamefont {E.}~\bibnamefont {Oelker}},\ and\ \bibinfo {author} {\bibfnamefont {J.}~\bibnamefont {Ye}},\ }\bibfield  {journal} {\bibinfo  {journal} {Physical Review X}\ }\textbf {\bibinfo {volume} {13}},\ \href {https://doi.org/10.1103/PhysRevX.13.041002} {10.1103/PhysRevX.13.041002} (\bibinfo {year} {2023})\BibitemShut {NoStop}%
\bibitem [{\citenamefont {Cripe}\ \emph {et~al.}(2018)\citenamefont {Cripe}, \citenamefont {Aggarwal}, \citenamefont {Singh}, \citenamefont {Lanza}, \citenamefont {Libson}, \citenamefont {Yap}, \citenamefont {Cole}, \citenamefont {McClelland}, \citenamefont {Mavalvala},\ and\ \citenamefont {Corbitt}}]{Cripe_RPL}%
  \BibitemOpen
  \bibfield  {author} {\bibinfo {author} {\bibfnamefont {J.}~\bibnamefont {Cripe}}, \bibinfo {author} {\bibfnamefont {N.}~\bibnamefont {Aggarwal}}, \bibinfo {author} {\bibfnamefont {R.}~\bibnamefont {Singh}}, \bibinfo {author} {\bibfnamefont {R.}~\bibnamefont {Lanza}}, \bibinfo {author} {\bibfnamefont {A.}~\bibnamefont {Libson}}, \bibinfo {author} {\bibfnamefont {M.~J.}\ \bibnamefont {Yap}}, \bibinfo {author} {\bibfnamefont {G.~D.}\ \bibnamefont {Cole}}, \bibinfo {author} {\bibfnamefont {D.~E.}\ \bibnamefont {McClelland}}, \bibinfo {author} {\bibfnamefont {N.}~\bibnamefont {Mavalvala}},\ and\ \bibinfo {author} {\bibfnamefont {T.}~\bibnamefont {Corbitt}},\ }\href {https://doi.org/10.1103/PhysRevA.97.013827} {\bibfield  {journal} {\bibinfo  {journal} {Physical Rev.iew A}\ }\textbf {\bibinfo {volume} {97}},\ \bibinfo {pages} {013827} (\bibinfo {year} {2018})}\BibitemShut {NoStop}%
\bibitem [{\citenamefont {Pagano}(2024)}]{thesis_Pagano}%
  \BibitemOpen
  \bibfield  {author} {\bibinfo {author} {\bibfnamefont {R.}~\bibnamefont {Pagano}},\ }\emph {\bibinfo {title} {Thermal Noise Measurement Below the Standard Quantum Limit}},\ \href {https://repository.lsu.edu/gradschool_dissertations/6651/} {\bibinfo {type} {Phd thesis}},\ \bibinfo  {school} {Louisiana State University} (\bibinfo {year} {2024})\BibitemShut {NoStop}%
\bibitem [{\citenamefont {Corbitt}\ \emph {et~al.}(2005)\citenamefont {Corbitt}, \citenamefont {Chen},\ and\ \citenamefont {Mavalvala}}]{Corbitt_mathematical}%
  \BibitemOpen
  \bibfield  {author} {\bibinfo {author} {\bibfnamefont {T.}~\bibnamefont {Corbitt}}, \bibinfo {author} {\bibfnamefont {Y.}~\bibnamefont {Chen}},\ and\ \bibinfo {author} {\bibfnamefont {N.}~\bibnamefont {Mavalvala}},\ }\href {https://doi.org/10.1103/PhysRevA.72.013818} {\bibfield  {journal} {\bibinfo  {journal} {Physical Review A}\ }\textbf {\bibinfo {volume} {72}},\ \bibinfo {pages} {013818} (\bibinfo {year} {2005})}\BibitemShut {NoStop}%
\bibitem [{\citenamefont {Vyatchanin}\ and\ \citenamefont {Zubova}(1995)}]{vyatchanin_quantum_variation_measurement}%
  \BibitemOpen
  \bibfield  {author} {\bibinfo {author} {\bibfnamefont {S.}~\bibnamefont {Vyatchanin}}\ and\ \bibinfo {author} {\bibfnamefont {E.}~\bibnamefont {Zubova}},\ }\href {https://doi.org/10.1016/0375-9601(95)00280-G} {\bibfield  {journal} {\bibinfo  {journal} {Physical Review A}\ }\textbf {\bibinfo {volume} {201}},\ \bibinfo {pages} {269} (\bibinfo {year} {1995})}\BibitemShut {NoStop}%
\bibitem [{\citenamefont {Chen}\ \emph {et~al.}(2011)\citenamefont {Chen}, \citenamefont {Danilishin}, \citenamefont {Khalili},\ and\ \citenamefont {Mueller-Ebhardt}}]{chen_qnd_for_gw_detectors}%
  \BibitemOpen
  \bibfield  {author} {\bibinfo {author} {\bibfnamefont {Y.}~\bibnamefont {Chen}}, \bibinfo {author} {\bibfnamefont {S.~L.}\ \bibnamefont {Danilishin}}, \bibinfo {author} {\bibfnamefont {F.~Y.}\ \bibnamefont {Khalili}},\ and\ \bibinfo {author} {\bibfnamefont {H.}~\bibnamefont {Mueller-Ebhardt}},\ }\href {https://doi.org/10.1007/s10714-010-1060-y} {\bibfield  {journal} {\bibinfo  {journal} {General Relativity And Gravitation}\ }\textbf {\bibinfo {volume} {43}},\ \bibinfo {pages} {671} (\bibinfo {year} {2011})}\BibitemShut {NoStop}%
\bibitem [{\citenamefont {Abbott}\ and\ \citenamefont {et. al.}(2022)}]{LIGO_GW_from_pulsars}%
  \BibitemOpen
  \bibfield  {author} {\bibinfo {author} {\bibfnamefont {R.}~\bibnamefont {Abbott}}\ and\ \bibinfo {author} {\bibnamefont {et. al.}},\ }\bibfield  {journal} {\bibinfo  {journal} {Astrophysical Journal}\ }\textbf {\bibinfo {volume} {935}},\ \href {https://doi.org/10.3847/1538-4357/ac6acf} {10.3847/1538-4357/ac6acf} (\bibinfo {year} {2022})\BibitemShut {NoStop}%
\bibitem [{\citenamefont {Abbott}\ and\ \citenamefont {et. al.}(2008)}]{LIGO_crab_pulsar}%
  \BibitemOpen
  \bibfield  {author} {\bibinfo {author} {\bibfnamefont {B.}~\bibnamefont {Abbott}}\ and\ \bibinfo {author} {\bibnamefont {et. al.}},\ }\href {https://doi.org/10.1086/591526} {\bibfield  {journal} {\bibinfo  {journal} {ASTROPHYSICAL JOURNAL LETTERS}\ }\textbf {\bibinfo {volume} {683}},\ \bibinfo {pages} {L45} (\bibinfo {year} {2008})}\BibitemShut {NoStop}%
\bibitem [{\citenamefont {Agazie}\ and\ \citenamefont {et. al.}(2023)}]{NANOGrav_15yr}%
  \BibitemOpen
  \bibfield  {author} {\bibinfo {author} {\bibfnamefont {G.}~\bibnamefont {Agazie}}\ and\ \bibinfo {author} {\bibnamefont {et. al.}},\ }\bibfield  {journal} {\bibinfo  {journal} {Astrophysical Journal Letters}\ }\textbf {\bibinfo {volume} {951}},\ \href {https://doi.org/10.3847/2041-8213/acdac6} {10.3847/2041-8213/acdac6} (\bibinfo {year} {2023})\BibitemShut {NoStop}%
\bibitem [{\citenamefont {Gardner}\ \emph {et~al.}(2025)\citenamefont {Gardner}, \citenamefont {Gefen}, \citenamefont {Haine}, \citenamefont {Hope}, \citenamefont {Preskill}, \citenamefont {Chen},\ and\ \citenamefont {McCuller}}]{gardner_stochastic_quantum_limit}%
  \BibitemOpen
  \bibfield  {author} {\bibinfo {author} {\bibfnamefont {J.~W.}\ \bibnamefont {Gardner}}, \bibinfo {author} {\bibfnamefont {T.}~\bibnamefont {Gefen}}, \bibinfo {author} {\bibfnamefont {S.~A.}\ \bibnamefont {Haine}}, \bibinfo {author} {\bibfnamefont {J.~J.}\ \bibnamefont {Hope}}, \bibinfo {author} {\bibfnamefont {J.}~\bibnamefont {Preskill}}, \bibinfo {author} {\bibfnamefont {Y.}~\bibnamefont {Chen}},\ and\ \bibinfo {author} {\bibfnamefont {L.}~\bibnamefont {McCuller}},\ }\href {https://doi.org/10.1103/h91r-4ws9} {\bibfield  {journal} {\bibinfo  {journal} {PRX Quantum}\ }\textbf {\bibinfo {volume} {6}},\ \bibinfo {pages} {030311} (\bibinfo {year} {2025})}\BibitemShut {NoStop}%
\end{thebibliography}%
\bibliographystyle{apsrev4-2}

\end{document}